\documentclass[%
 aip,
 amsmath,amssymb,
 reprint,
]{revtex4-1}

\usepackage{graphicx}
\usepackage{dcolumn}
\usepackage{bm}
\usepackage[utf8]{inputenc}
\usepackage[T1]{fontenc}
\usepackage{mathptmx}
\usepackage{hyperref}

\begin{document}

\title{Muon Imaging of Hydrotreatment Reactors}
\author{R.A.~Martínez-Rivero}
\email{rafael2248058@correo.uis.edu.co}
\affiliation{Escuela de Física, Universidad Industrial de Santander, Bucaramanga-Colombia.}
\affiliation{Departamento de Física, Universidad Simón Bolivar, Caracas-Venezuela.}
\author{C.~Sarmiento-Cano}
\email{christian.sarmiento@correo.uis.edu.co}
\affiliation{Escuela de Física, Universidad Industrial de Santander, Bucaramanga-Colombia.}
\author{D.~Castillo-Morales}
\affiliation{EcoNuclear S.A.S., Bucaramanga-Colombia.}
\author{J.~Perea-Pérez}
\affiliation{Escuela de Ingenieria en Sistemas, Universidad Industrial de Santander, Bucaramanga-Colombia.}
\author{V.G.~Baldovino-Medrano}
\email{vicbaldo@uis.edu.co}
\affiliation{Centro de Investigaciones en Catálisis,  Universidad Industrial de Santander, Bucaramanga-Colombia.}
\affiliation{Escuela de Ingenieria Química, Universidad Industrial de Santander, Bucaramanga-Colombia.}
\affiliation{Laboratorio de Ciencia de Superficies, Universidad Industrial de Santander, Bucaramanga-Colombia.}
\author{J.D.~Sanabria-Gómez}
\email{jsanabri@uis.edu.co}
\affiliation{Escuela de Física, Universidad Industrial de Santander, Bucaramanga-Colombia.}
\author{L.A.~Núñez}
\email{lnunez@uis.edu.co}
\affiliation{Escuela de Física, Universidad Industrial de Santander, Bucaramanga-Colombia.}
\affiliation{Departamento de Física, Universidad de Los Andes, Mérida-Venezuela.}
\date{\today}

\begin{abstract}
This study presents the design and simulation-based validation of a muon imaging system tailored for potential applications in industrial hydrotreatment units. The system is built around a two-panel plastic scintillator hodoscope, equipped with silicon photomultipliers and read-out via a CAEN FERS-A5202 acquisition system. The detector was calibrated using a stepwise “staircase” method and characterized under open-sky and controlled conditions. We conducted muon flux attenuation measurements to validate its response using variable lead shielding. We agreed with simulations generated using the MEIGA framework and realistic cosmic ray spectra from the ARTI simulation chain. With the detector response validated, we modelled muon transmission through a realistic 3D representation of a benchmark unit for producing clean fuels, incorporating internal variations in the density of the fixed bed inside the unit. By reconstructing angular muon fluxes and computing relative attenuation maps, we demonstrated the system’s capability to detect internal density contrasts. Simulation results indicate that 20 hours of exposure to vertical muon flux is sufficient to retrieve structural information. In comparison, inclined configurations (30° and 60° from vertical) require extended exposure times—up to 8 days—yet remain feasible within industrial monitoring schedules.
These findings highlight the feasibility of muography as a non-invasive diagnostic tool for complex industrial infrastructure. The proposed system shows strong potential for real-time monitoring of catalyst bed integrity and long-term structural analysis in high-pressure chemical reactors.
\end{abstract}

\maketitle

\section{Introduction}
Muon radiography -also known as muon tomography or simply muography- is a non-invasive imaging technique that uses naturally occurring cosmic-ray muons to explore the internal structure of large and dense objects. Muons are highly penetrating, charged particles produced in the upper atmosphere when high-energy cosmic rays collide with air nuclei. These interactions generate mesons, such as charged pions and kaons, which decay into muons via weak interaction processes. Due to their relatively high mass, approximately $200$ times that of the electron, and their broad energy spectrum, muons interact weakly with matter. They can travel through hundreds of meters of rock, metal, or concrete before being absorbed or deflected. Despite gradual attenuation, this penetrating power makes muons uniquely suitable for probing deep within otherwise inaccessible structures~\cite{GaisserEngelResconi2016}.

In contrast to X-rays or gamma rays, which require artificial sources and are limited by their low penetration depth in dense materials, muons provide a naturally abundant and highly penetrating alternative for radiographic imaging. In a typical muography setup, a detector is placed behind or beneath the object of interest to measure the angular distribution of incoming muons. By comparing the observed muon flux to the expected flux under open-sky conditions, one can infer the integrated density (or opacity) along each line of sight. Denser regions attenuate the muon flux more effectively, producing detectable shadows or deficits corresponding to structural anomalies. This approach enables the reconstruction of two- or three-dimensional density maps capable of revealing voids, dense inclusions, or other internal features. Depending on the detector configuration and exposure time, muography can achieve spatial resolutions ranging from tens of centimetres to several meters, making it a powerful tool for applications in geophysics, civil engineering, archaeology, and industrial diagnostics~\citep{TanakaEtal2023}.

Initially developed for geoscientific applications-- such as imaging the internal structure of volcanoes~\cite{TanakaEtal2007b}-- muography has advanced considerably thanks to innovations in detector technology, simulation frameworks, and data analysis algorithms~\cite{PezzottiEtal2025}. Modern muon detectors, including scintillator-based hodoscopes coupled with silicon photomultipliers (SiPMs), provide compact, durable, and energy-efficient solutions that are well-suited for deployment in both remote and industrial environments~\cite{PenarodriguezEtal2020, CalderonEtal2020, Tishevsky2024_SPD}. In parallel, the development of simulation frameworks such as ARTI~\cite{SarmientoCano_2022}, MEIGA~\cite{TaboadaEtal2022} and MUYSC~\cite{PenaRodriguezEtal2024} has enabled the realistic modelling of cosmic ray fluxes, particle interactions, and detector response under a wide range of environmental and operational conditions. 

In the oil and gas industry, muography offers a powerful non-invasive alternative for inspecting large-scale, high-pressure process units. Equipment such as hydrotreatment reactor towers, hydrocrackers, and catalytic reformers operate under extreme thermal and mechanical stresses, which pose challenges for traditional inspection techniques that are often intrusive, hazardous, or insufficiently sensitive. Muography can detect internal density anomalies --such as catalyst bed compaction, fouling, or mechanical deformation-- by analysing the angular attenuation of cosmic muons as they pass through the reactor vessel. These density variations frequently correlate with loss of reactor performance or precursors to structural failure, making early detection vital for maintaining operational safety, optimising performance, and reducing downtime~\cite{Furimsky_1999, Argyle_2015, Duarte_2019}.

This study focuses on a muon imaging system developed for hydrotreatment reactors, which is a critical component in clean fuels complying with tight environmental legislation regarding sulfur, nitrogen, aromatics, and other hamful pollutants that would otherwise be emitted to the atmosphere hence aggravating the current environmental crisis. Over time, the catalyst beds inside these reactors degrade due to chemical poisoning, sintering, or the accumulation of contaminants. Traditional inspection methods often require shutdowns or physical access to the reactor interior, posing logistical and safety challenges. In contrast, muography enables in situ imaging of internal changes without interrupting operations. Our system can resolve density contrasts as tiny as a few g/cm³—validated through simulations and experimental shielding tests—making it suitable for detecting phenomena such as catalyst settling or uneven compaction, with practical resolution and manageable exposure times.

This paper is organised as follows. Section II outlines the methodology used to estimate the ground-level cosmic muon flux, leveraging the ARTI simulation framework. Section III describes the muon detector, including its mechanical design, data acquisition system, and calibration procedure. Section IV details the simulation environment developed in MEIGA to evaluate detector performance under realistic cosmic-ray conditions. Section V presents the experimental validation of the system, including measurements under open-sky and controlled environments and attenuation tests using lead shielding. Section VI introduces the hydrotreatment unit model, describes its internal composition and geometry, and presents the results of the muographic imaging simulations. Finally, Section VII summarises the conclusions and evaluates the feasibility of deploying muography as a real-time diagnostic tool in industrial reactors.

\section{Cosmic muon component at the Ground Level}
To estimate the secondary particle flux at ground level and simulate the muon flux at specific observation sites, we used the \textsc{ARTI} framework\footnote{\url{https://github.com/lagoproject/arti}}~\cite{SarmientoCano_2022}. \textsc{ARTI} is a comprehensive simulation toolkit that models the generation and propagation of secondary particles resulting from the interaction of primary cosmic rays with Earth's atmosphere. It supports location-specific simulations by incorporating geographic parameters (latitude, longitude, and altitude), real-time atmospheric profiles, geomagnetic conditions, and detector configurations. The modular architecture of \textsc{ARTI}—built using a combination of \texttt{C++}, \texttt{Fortran}, \texttt{Bash}, and \texttt{Perl}—ensures smooth integration with widely used astroparticle simulation tools.

Our simulations follow the methodology outlined in~\cite{AsoreyNunezSuarez2018, Asorey_etal2018}. In this framework, the primary cosmic ray flux $\Phi$, injected at the top of the atmosphere (at an altitude of $112$~km above sea level), is parameterised as:
\begin{equation}
  \Phi(E_p, Z, A, \Omega) \simeq j_0(Z,A)
  \left(\frac{E_p}{E_0}\right)^{\alpha(E_p,Z,A)}\, .
\end{equation}
Here, $E_p$ denotes the energy of the primary cosmic ray particle, and $\alpha(E_p, Z, A)$ is the spectral index, which is treated as constant over the relevant energy range (i.e., $\alpha \equiv \alpha(Z, A)$) from a few GeV up to $10^6$~GeV. Each galactic cosmic ray (GCR) species is identified by its atomic number $Z$ and the mass number $A$, while $j_0(Z, A)$ represents the measured flux at the top of the atmosphere for reference energy $E_0 = 10^3$, GeV.

The \textsc{ARTI} simulation chain incorporates the CORSIKA code~\cite{CORSIKA}. CORSIKA simulates the interaction of each primary GCR with atmospheric nuclei and tracks the resulting secondary particles down to their minimum energy thresholds. These thresholds are particle-dependent: $E_s \geq 5$~MeV for muons and hadrons (excluding pions), and $E_s \geq 5$~keV for electrons, pions, and gamma photons.

Because the atmospheric density profile strongly influences the development of air showers, CORSIKA requires accurate atmospheric input. For this purpose, we employed MODTRAN-based models~\cite{Kneizys1996, Grisalescasadiegos2022}, which provide location-specific atmospheric profiles. A detailed description of the methodology can be found in~\cite{SarmientoCano_2022} and references therein.

\begin{figure}[h!]
    \centering
    \includegraphics[width=0.45\textwidth]{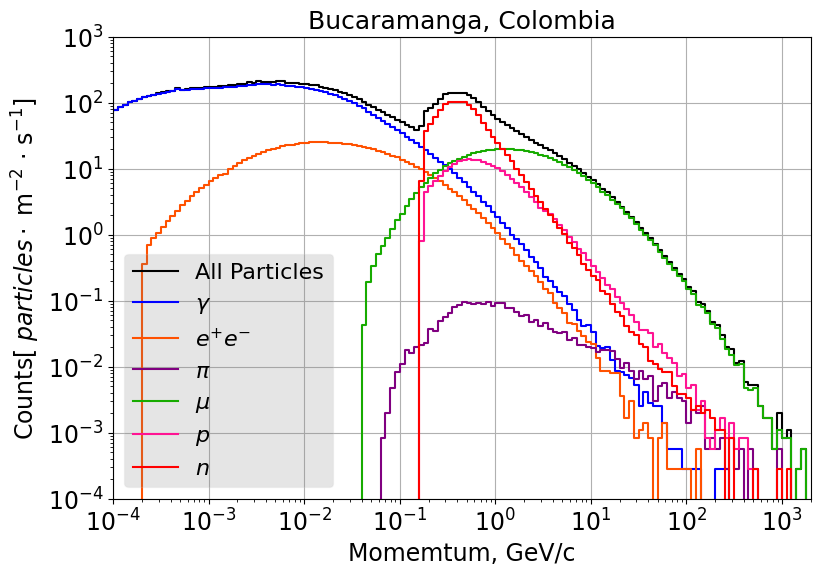}
    \caption{Energy spectrum of secondary particles at Bucaramanga, Colombia (956\,m a.s.l.). The total flux of particles reaching the ground is shown in black. The electromagnetic component includes gamma photons (blue), electrons, and positrons (orange). The proton component is represented in purple, while the muon component is shown in green.
}
    \label{fig:arti_spectrum}
\end{figure}

Figure~\ref{fig:arti_spectrum} shows the simulated energy spectrum of secondary particles produced by cosmic ray showers over Bucaramanga, Colombia, as obtained using the \textsc{ARTI} framework~\cite{SarmientoCano_2022}. The spectrum includes contributions from various particle species: photons ($\gamma$), electrons and positrons ($e^+e^-$), pions ($\pi$), muons ($\mu$), protons ($p$), and neutrons ($n$).

Among these, muons are particularly relevant for muography applications. They are mainly produced within the energy range of 0.1 to $100$~GeV/$c$, and their flux dominates at higher momenta due to their long atmospheric attenuation length. In contrast, electrons and positrons are generated at lower energies—typically between $10^{-3}$ and $1$~GeV/$c$—but they experience significant attenuation before reaching the ground, primarily due to electromagnetic interactions that cause rapid energy loss during atmospheric propagation.

Overall, the energy spectrum of secondary particles exhibits a characteristic power-law behaviour shaped by the interplay of energy loss, scattering, and particle interaction processes occurring throughout the atmosphere.

\begin{figure*}[ht!]
    \centering    
    \includegraphics[width=0.4\textwidth]{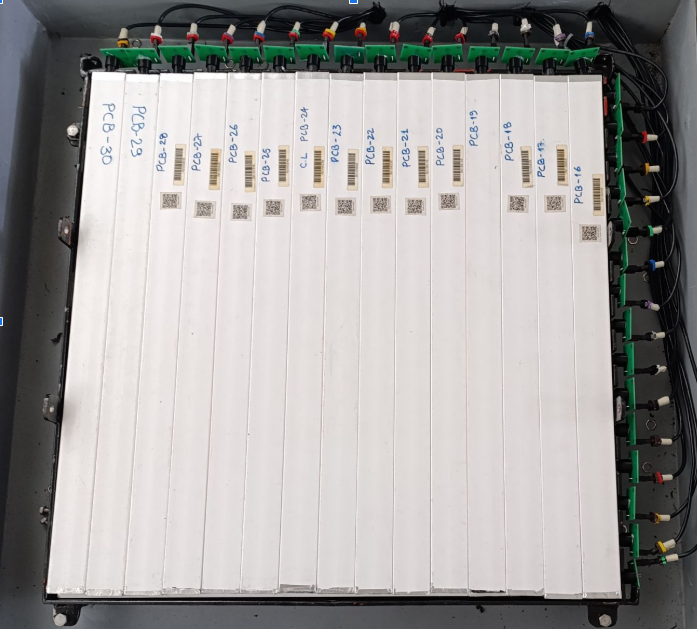}
    \includegraphics[width=0.5\textwidth]{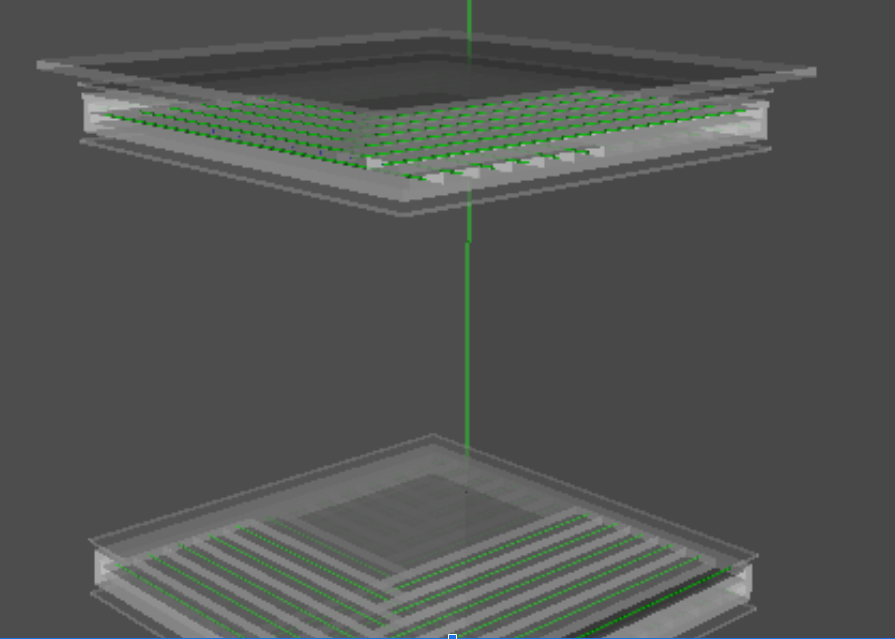}
    \caption{(a) Photograph of the experimental muon detector setup. The system comprises two orthogonal detection planes, each comprising 15 plastic scintillator bars arranged to form a $15 \times 15$ matrix. Each scintillator bar is optically coupled to a printed circuit board, which interfaces with the front-end readout electronics. Scintillation light produced by traversing muons is captured by wavelength-shifting optical fibres and directed toward silicon photomultipliers. These sensors are connected to a CAEN FERS-A5202 acquisition module, which processes the signals for event reconstruction. The wiring and connectors ensure reliable signal transmission from the SiPMs to the data acquisition system, allowing for accurate reconstruction of cosmic muon trajectories.
(b) Simulated representation of the detector geometry as implemented in the MEIGA framework.
} 
    \label{fig:Detector}
\end{figure*}

\section{The Muon Telescope}
\subsection{The detector}
The detector employed in this study is a two-panel scintillator-based hodoscope specifically designed to detect and reconstruct the trajectories of atmospheric muons. Each detection panel consists of a $15 \times 15$ matrix of plastic scintillator strips, each measuring $60\text{cm} \times 4\text{cm} \times 1\text{cm}$. The scintillators are fabricated from polystyrene (Dow Styron 663) and are externally coated with titanium dioxide ($\text{TiO}_2$) to enhance internal light reflection. The scintillator composition is doped with $1$\% PPO and $0.03$\% POPOP, yielding an absorption cutoff near $40$nm and a peak emission at $420$nm.

A multi-clad wavelength-shifting fibre (Saint-Gobain BCF-92) is embedded within each scintillator strip, efficiently capturing and guiding the scintillation photons to the photodetector. The photodetectors are silicon photomultipliers (S13360-1350CS by Hamamatsu), each with an active area of $1.3 \times 1.3\text{mm}^2$. These SiPMs feature $667$ microcells, with a fill factor of $74$\% and a photon detection efficiency of approximately $40$\% at $450$nm. The gain of each device can be tuned in the range of $10^5$ to $10^6$~(see references \cite{VasquezEtal2020,PenarodriguezEtal2020}).

The hodoscope operates in a coincidence mode between the two parallel panels, enabling the identification of muon passage events and the reconstruction of their angular trajectories. The system's spatial resolution is primarily governed by the dimensions of the scintillator strips and the time resolution of the SiPM signal readout.

Figure~\ref{fig:Detector}(a) shows a picture of the assembled detector, including the two scintillator planes, the signal-processing printed circuit boards, and the associated wiring for data acquisition. This representation highlights the mechanical configuration and component integration of the system.

\subsection{The electronics}
\begin{figure}[h!]
    \centering
    \includegraphics[width=0.45\textwidth]{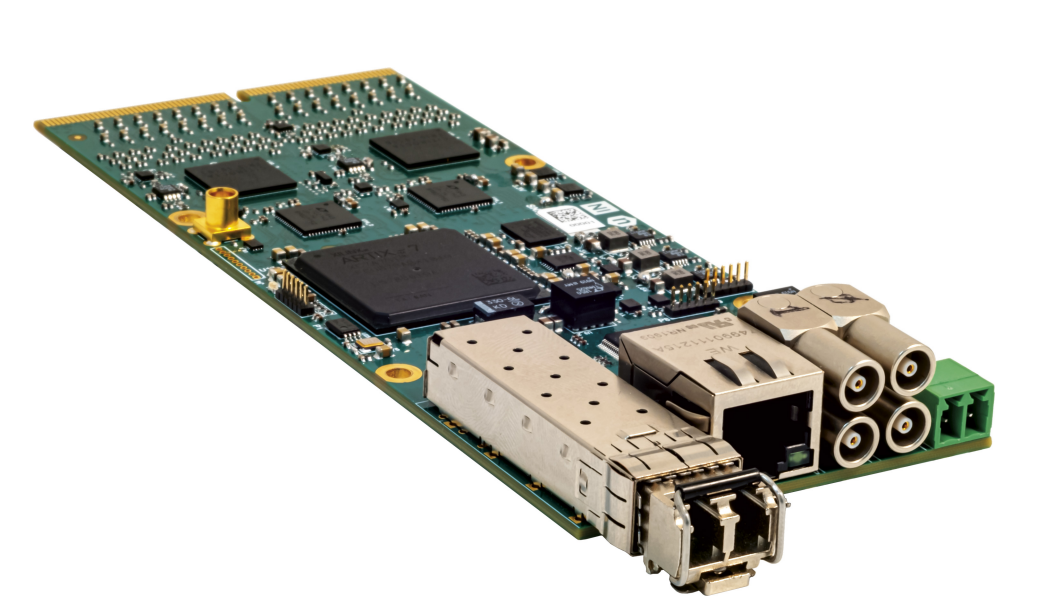}
    \caption{General view of the 64-channel FERS-A5202 unit for SiPM readout naked version. \cite{Venaruzzo2020}}
    \label{fig:FERS}  
\end{figure}

The data acquisition system employed in this study is built around the CAEN FERS-A5202 platform, a front-end readout solution designed explicitly for large-scale detector arrays. This system supports a wide range of sensor technologies, including SiPMs, multi-anode photomultiplier tubes, silicon strip detectors, wire chambers, gas electron multipliers, and gas-filled detection tubes~\cite{Tishevsky2024, Tishevsky2024_SPD, Zakharov2024}. An overview of the FERS-A5202 hardware architecture, including its key components and interface ports, is shown in Figure~\ref{fig:FERS} and further detailed in references~\cite{FERS5200, UM7945}.

At the heart of the A5202 module are two Citiroc-1A ASICs, each providing 32 channels for 64 independent readout lines. These ASICs are tailored for high-density, high-resolution applications that demand fast timing and precise pulse analysis. The system operates in a trigger-less acquisition mode, enabling continuous real-time data collection without external trigger signals. Each readout channel includes a charge-sensitive preamplifier, a fast discriminator, and dual-gain signal processing paths, allowing for an extended dynamic range. Furthermore, each channel is equipped with an 8-bit digital-to-analog converter (DAC), allowing precise adjustment of the SiPM bias voltage within a $20$V–$85$V range. A built-in temperature feedback loop compensates for gain drift, ensuring stable performance during prolonged muography campaigns, which is particularly important in oil industry settings where environmental conditions may vary.

One of the most notable features of the FERS-A5202 system is its capability to perform both Time of Arrival (ToA) and Time over Threshold (ToT) measurements. These features allow for high-precision timing and accurate characterisation of pulse amplitudes, essential for reconstructing muon trajectories with high spatial resolution. Although the Citiroc-1A does not include a native time-to-digital converter (TDC), this functionality is implemented within the FPGA of the companion DT5202 module, achieving a time resolution of better than $100$~ps RMS through a $0.5$~ns-resolution TDC. Such precision is critical for resolving fine-grained structural details in subsurface muon imaging.

The FERS-A5202 also supports multiple acquisition modes tailored to diverse experimental requirements. Among these is the Spectroscopy Mode, which enables pulse height analysis to determine energy deposition. This mode supports applications such as muon energy discrimination and material identification—key elements when applying muography in the oil industry.

For data management and user interaction, the FERS-A5202 supports various communication protocols—including USB, Ethernet, and TDlink (optical fibre)—ensuring seamless integration with scalable data acquisition  architectures. User-friendly software tools like Janus provide an intuitive graphical interface for system configuration and real-time histogram visualisation. The FersRun framework—built on the ROOT platform—enables sophisticated event filtering, data processing, and analysis for more advanced operations.

A key strength of the FERS architecture lies in its scalability. The modular design expands the system to handle up to 8192 acquisition channels, ideally suited for large-area muography detectors. This capability is particularly beneficial for industrial monitoring applications, such as imaging the internal structure of catalytic reforming towers, hydrocrackers, or other large-scale equipment in oil refineries.
We used the Janus software \cite{UM7946} to configure and control the data acquisition system. Janus supports multiple acquisition modes—including spectroscopy, counting, and timing—and offers an intuitive graphical interface for real-time monitoring and parameter configuration. During data collection, output files were stored in ASCII and binary formats to enable flexible post-processing and long-term archiving. For advanced data handling, we employed the FersRun framework, a ROOT-based tool that facilitates real-time data visualisation, signal filtering, and correlation analysis. This software combination significantly streamlined the acquisition workflow and simplified the interpretation of the large datasets produced during muography experiments.

\subsection{The calibration}
We implemented a detailed calibration procedure based on the staircase method to ensure uniform performance across all detection channels. The calibration focused on optimising two key parameters: the bias voltage applied to each SiPM and the Time Discriminator (TD) threshold used for event detection. Using the CAEN FERS-A5202 module, we adjusted the bias voltage individually for each channel via its integrated 8-bit DACs. This allowed compensation for variations in the breakdown voltage between SiPMs, ensuring consistent gain and stable performance across the entire detector array.

The calibration process began with a bias voltage scan. For each SiPM, we gradually increased the voltage while measuring gain and noise levels using a pulsed LED source. This step helped us determine the optimal operating point that balanced signal amplification with noise suppression, maximising the signal-to-noise ratio.

Subsequently, we calibrated the TD threshold to suppress dark noise. To isolate this effect, we temporarily disconnected the scintillator bars from the SiPMs and measured only the intrinsic dark count rate. By progressively increasing the TD threshold and monitoring the corresponding count rate, we identified the point at which dark noise was significantly reduced without compromising the system's sensitivity to actual photon events. This approach ensured effective noise rejection while maintaining high detection efficiency—which is critical for accurate muon trajectory reconstruction.
\begin{figure}[hb!]
    \centering
    \includegraphics[width=1.1\linewidth]{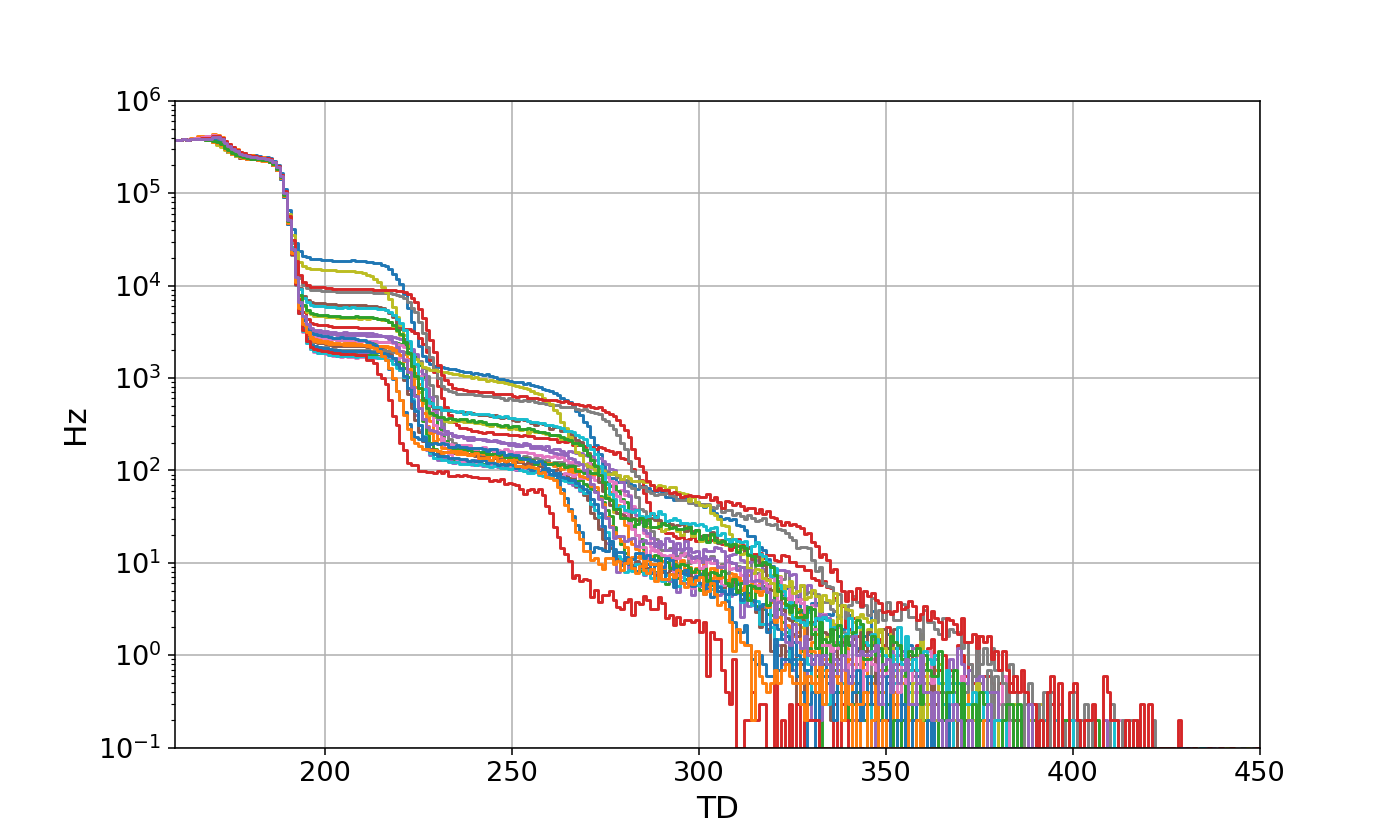}
    \caption{Staircase scan results for the 30 SiPM channels in Panel 2. The count rate (Hz) is plotted as a function of the TD threshold. A clear decrease in the count rate is observed as the TD increases, corresponding to the suppression of dark counts. The chosen operating point was TD = 420, ensuring an optimal balance between noise suppression and signal efficiency.}
    \label{fig:staircase_scan}
\end{figure}
To calibrate the SiPMs, we fixed the bias voltage at $56$~V and adjusted the TD threshold to $420$. This value was selected to strike a balance between minimising dark count rates and preserving sensitivity to genuine photon signals. To validate and refine this setting, we employed the staircase scan method across all $60$ independent detection channels.

During the scan, the TD threshold was incrementally increased in discrete steps. We recorded the corresponding count rate at each step, allowing us to evaluate how the system responded to different threshold levels.
The results for $30$ channels in Panel $2$ are shown in Figure~\ref{fig:staircase_scan}(b). As expected, the count rate consistently decreased with increasing TD threshold, effectively suppressing dark noise and electronic fluctuations. The response curves confirm that a threshold of 420 provides an optimal compromise—sufficiently high to reject spurious counts while maintaining high detection efficiency for actual photon events.

\section{Detector Simulation}
This study employs the MEIGA framework~\cite{TaboadaEtal2022} to simulate the detector's response and assess its suitability for muography applications. MEIGA is a specialised simulation environment designed for cosmic ray studies. It integrates, into a unified and cohesive workflow, three essential components: cosmic-ray flux generation, particle propagation through matter, and detector respons. MEIGA offers flexible interfaces for defining detector geometry, controlling simulation parameters, and adapting the model to various experimental configurations.

This work implements the detector geometry in Geant4 using simple yet representative volumes for each significant component: scintillator bars, optical fibres, and structural supports. Rotation matrices, which position each component according to the physical detector layout, ensure the accurate spatial arrangement of these elements.

The scintillator response is modelled by simulating photon emission in the $2.00$~eV–$4.20$~eV energy range, covering both the visible and near-ultraviolet spectrum. This range corresponds to the photons produced during scintillation and subsequently detected by the optical system. The scintillator material is assigned a fixed refractive index of 1.5, which governs the speed and angle of photon propagation. Photon absorption lengths vary from $4$~cm to $24$~cm, with lower-energy photons being absorbed more readily. To accurately reflect temporal characteristics, both fast and slow scintillation components are included in the model, with the fast component dominating at higher photon energies.

Scintillation photons are transported to the detection system via optical fibres, which operate under the principle of total internal reflection. These fibres have a refractive index of $1.60$, facilitating efficient light transmission. However, absorption within the fibres is energy-dependent—higher-energy photons experience more significant attenuation losses during transport.

Finally, the simulation includes a comprehensive model of the photon detection system, with a particular focus on SiPMs. This model incorporates key optical properties, such as surface refractive index and quantum efficiency, which are defined as photon energy functions. These parameters are critical for accurately estimating photon detection probability and ensuring optimal performance of the complete detection system under realistic operational conditions.
\begin{figure*}[ht!]
    \centering
    \includegraphics[width=0.31\textwidth]{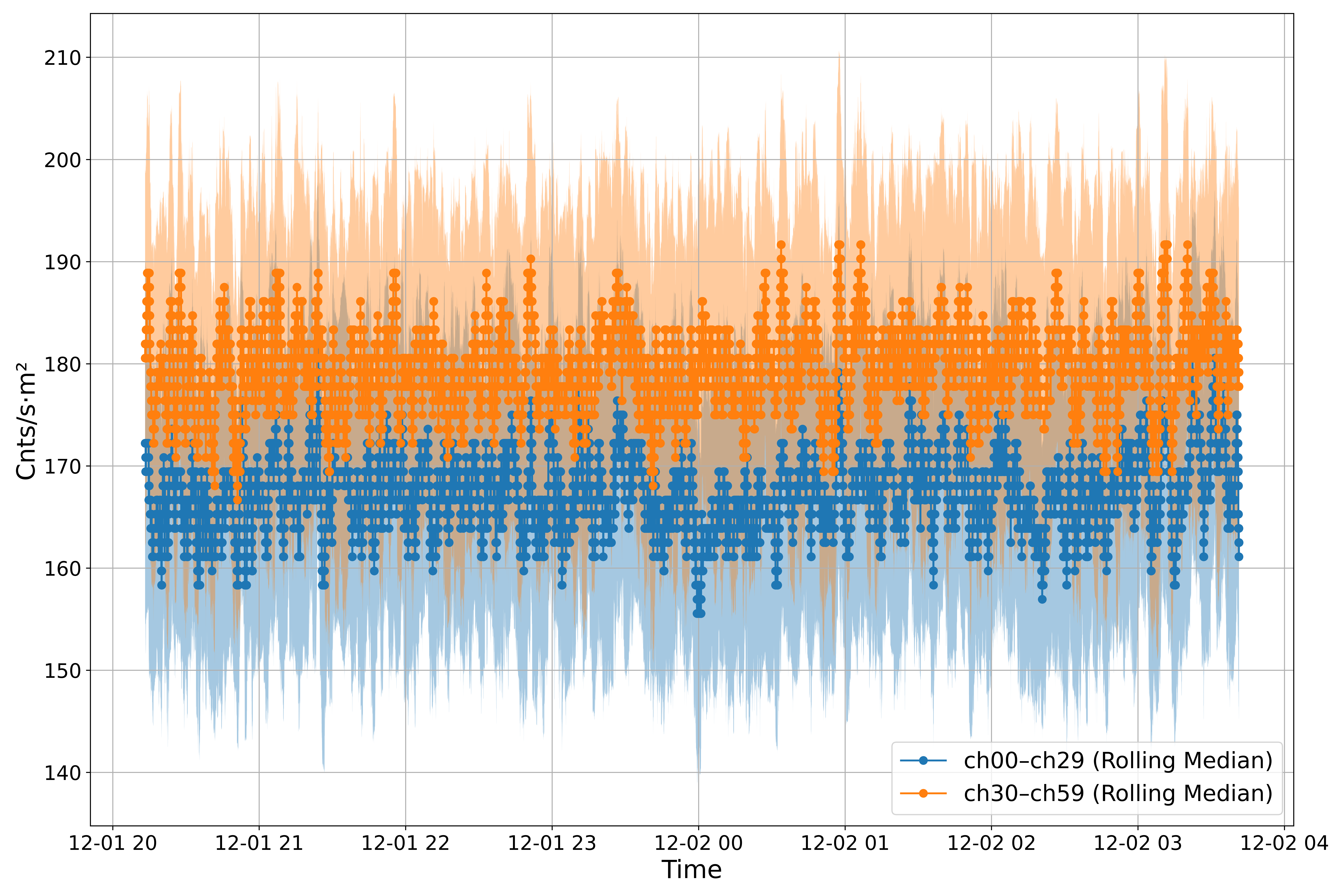}
    \includegraphics[width=0.31\textwidth]{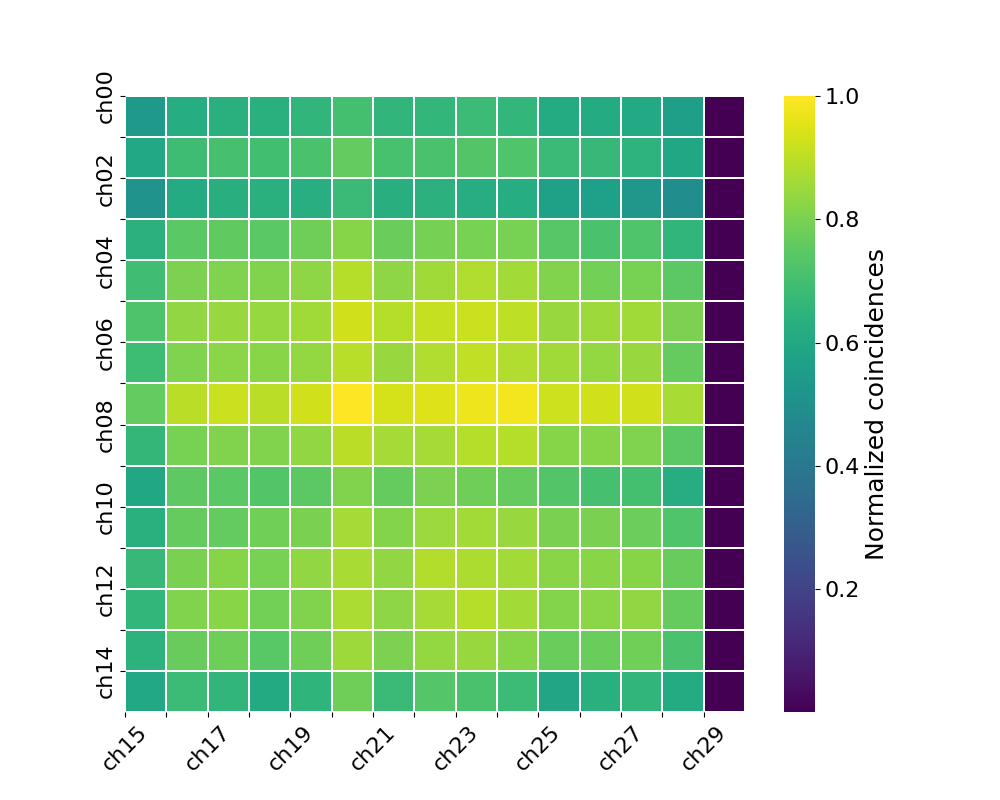}
    \includegraphics[width=0.31\textwidth]{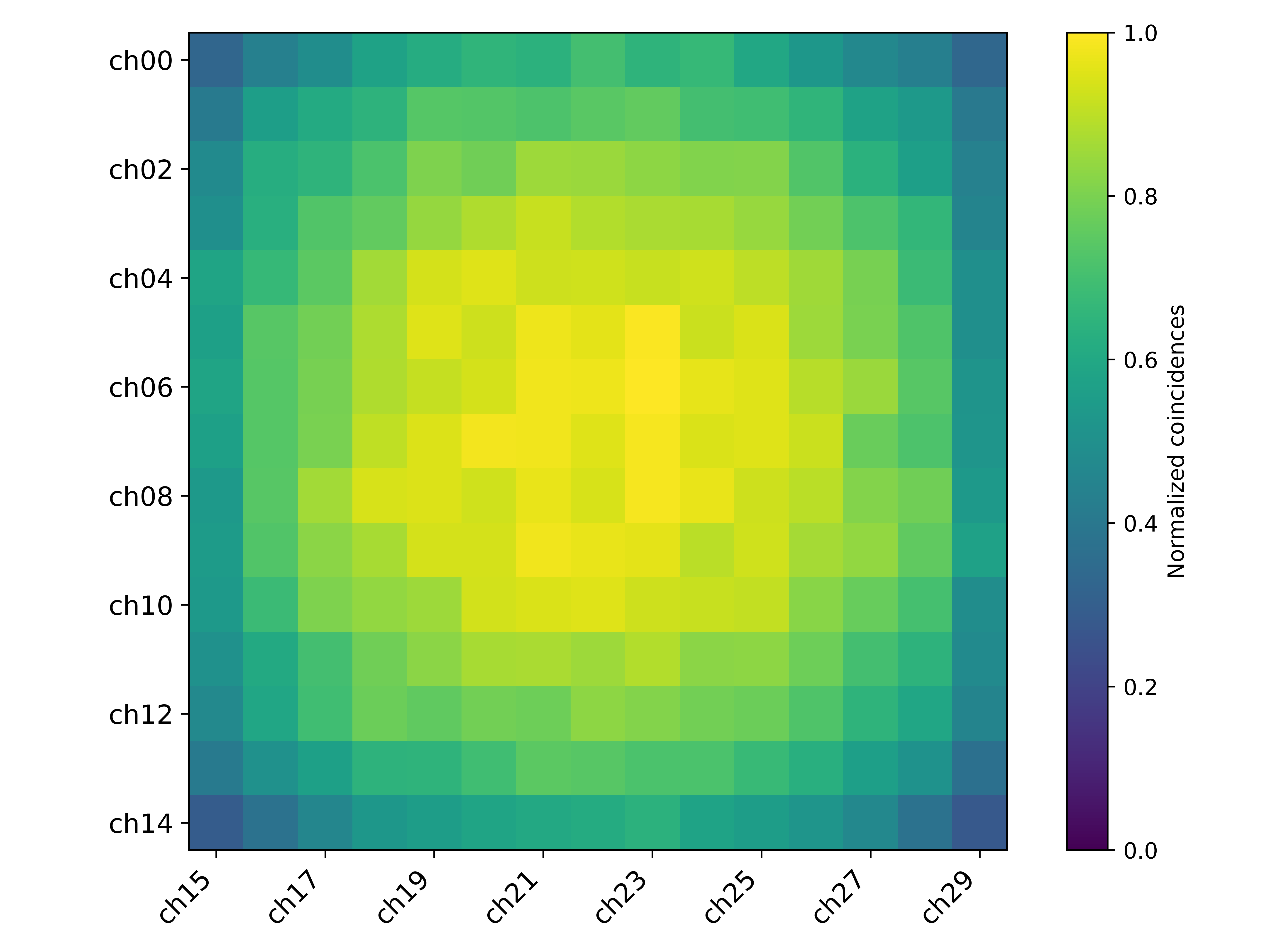}
    \caption{Results under open-sky conditions: (a) The time series of the detected flux for both panels demonstrates stable performance over time, indicating good temporal consistency. (b) Hit distribution in Panel 1 as measured by the 30 SiPM channels  are presented, with the final bar on the right intentionally disconnected. (c) A simulated hit distribution for Panel 1 under similar conditions is also shown. These plots highlight the substantial agreement between experimental measurements and simulations, confirming the system's stability and reliability even under uncontrolled outdoor environmental conditions.}
    \label{fig:comparison}
\end{figure*}
The interaction of light with detector surfaces plays a critical role in determining the system's overall efficiency. One key factor is surface roughness, which is quantified by a dedicated parameter and directly affects how photons scatter at the interface between the scintillator and optical fibres. We use a specular lobe reflection model to accurately model this effect, in which 20\% of incident photons are reflected within a narrow angular range. Pure specular reflections are intentionally excluded better to represent the behaviour of a realistically rough surface. A visual representation of the detector geometry, as implemented in the MEIGA framework, is provided in Figure~\ref{fig:Detector}, panel (b).
\begin{figure}[ht!]
    \centering
    \includegraphics[width=0.4\textwidth]{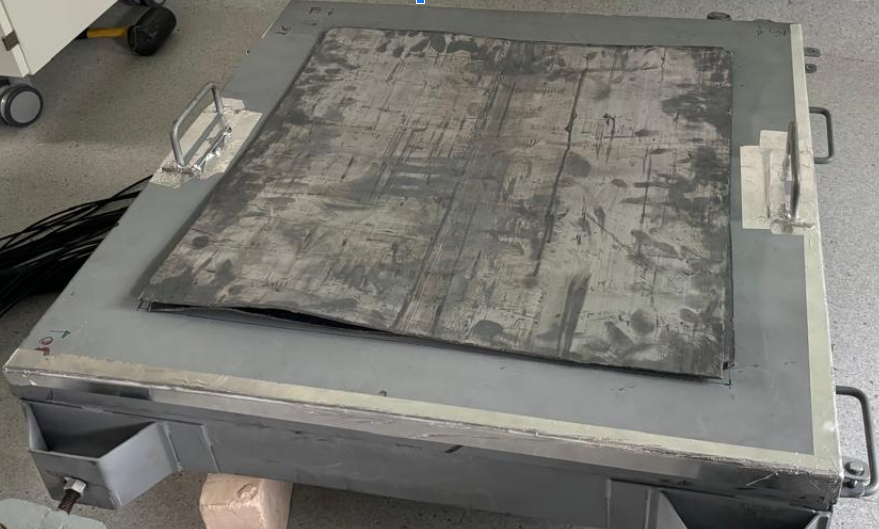}
    \caption{Photograph of the experimental setup showing lead sheets placed on top of the detector for shielding tests. The lead thickness was progressively increased to analyze its effect on muon detection.}
    \label{fig:pb_setup}
\end{figure}

\section{The simulation validation}
We conducted a series of simulations and experimental measurements under controlled conditions to validate the detector's response and ability to suppress background noise from soft particles. These tests focused on evaluating the attenuation effects introduced by lead shielding and analysing the performance of individual pixels within the hodoscope.
Simulations were carried out using the MEIGA framework, which incorporates secondary cosmic ray fluxes generated by the ARTI simulation chain. As previously described, the detector model included detailed geometric and physical configurations to replicate realistic operational conditions.

The cosmic ray flux was injected $1$~cm above the detector in the simulation setup. The injection area was defined as a circular region with a total area of $1$~m$^2$, ensuring a uniform distribution of incoming particles. The positions of incident cosmic ray particles were randomised within this area to preserve the natural stochastic nature of their arrival.

Experimental measurements performed under open-sky conditions yielded particle fluxes of approximately $167~\pm~16$~particles/m²/s for Panel 1 and $179~\pm~17$ particles/m²/s for Panel 2. Figure~\ref{fig:comparison}(a) displays the time series of real-time flux measurements for both panels throughout the exposure period, demonstrating stable performance and no significant correlation with ambient light variations, such as sunrise, thereby confirming the system's resistance to external light leakage.

Additionally, Figures~\ref{fig:comparison}(b) and (c) show the pixel coincidence maps for Panel 1 when triggered by Panel 2, comparing experimental and simulated data. Both maps exhibit a central concentration of events consistent with the expected acceptance pattern, validating the detector's angular response and confirming the accuracy of the simulation.

Simulations were performed to assess the effect of lead shielding on particle detection efficiency across various thicknesses: $0$~mm, $5$~mm, $10$~mm, $15$~mm, $20$~mm, $25$~mm, and $30$~mm. The detector was exposed to simulated cosmic ray flux for one hour for each configuration. To enhance the statistical reliability of the results, each scenario was repeated ten times using a bootstrap resampling technique, enabling a more accurate estimation of statistical fluctuations.

We performed experimental measurements using the actual detector setup under well-controlled conditions to complement the simulation results. These tests were carried out in a light-tight enclosure with regulated temperature, minimising environmental noise and ensuring stable detector performance throughout the acquisition period. The same set of lead shielding thicknesses, was employed. For each thickness level, data were acquired independently for one hour per detection panel, allowing us to isolate and assess the effect of incremental shielding on the measured particle flux. A photograph of the experimental arrangement, including the placement of lead sheets atop the detector, is shown in Figure~\ref{fig:pb_setup}.

After completing both simulations and experimental measurements, we conducted a comparative analysis by evaluating coincidence rates between specific scintillator bars. In particular, we selected a pair of aligned bars—one in each panel—and analysed the coincidence counts as a function of lead thickness. This approach enabled us to quantify the attenuation effect of the shielding material and assess the detector's ability to distinguish muons from soft background particles such as low-energy electrons and photons. The observed trends confirmed the system's sensitivity to shielding variations and its capability to suppress non-muon background, a critical feature for muography applications in environments with complex radiation fields—such as industrial or subsurface geological settings.
\begin{figure}
    \centering
    \includegraphics[width=0.42\textwidth]{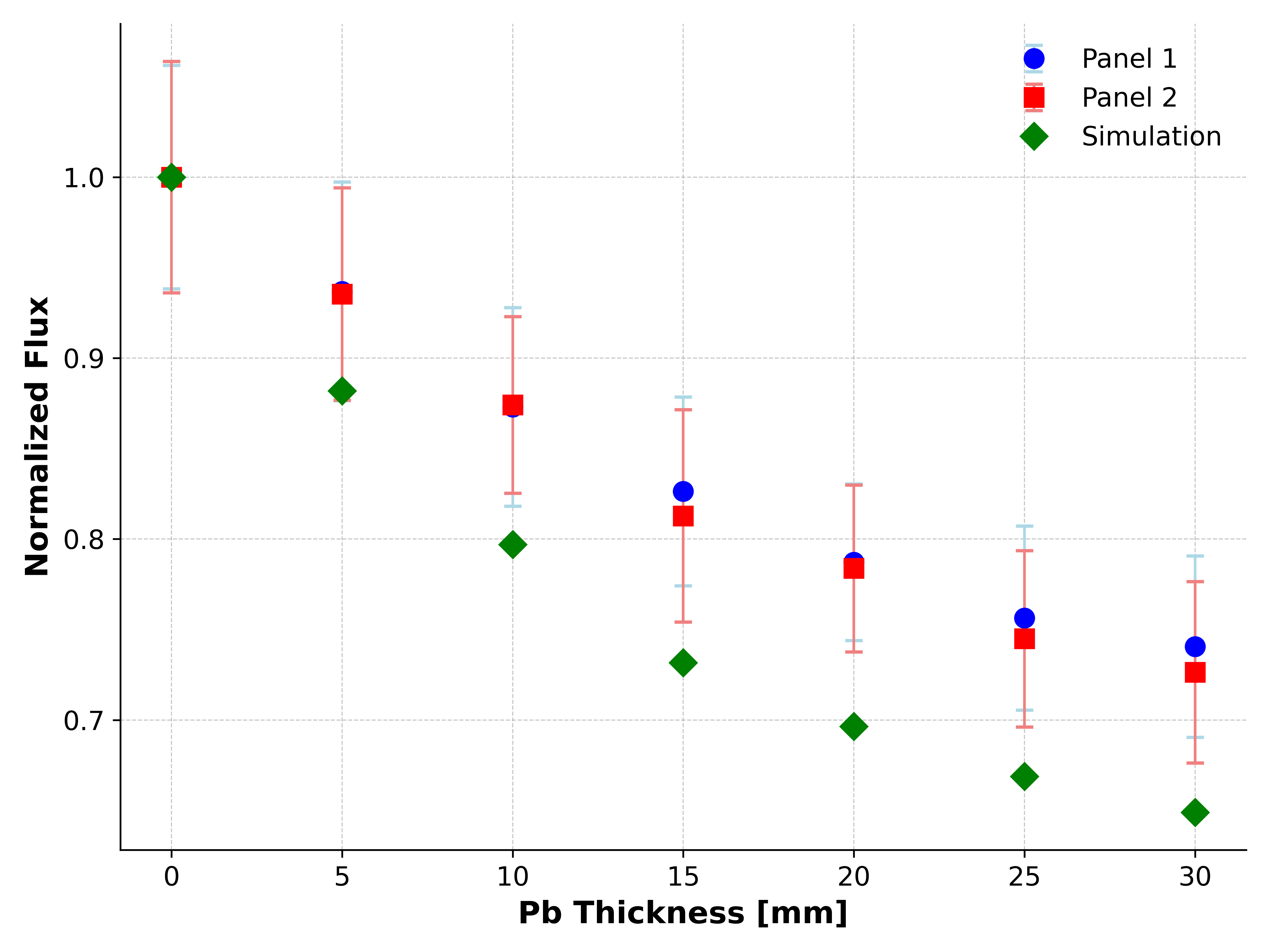}
    \caption{Normalized flux measured for both detection panels as a function of lead thickness. A clear attenuation trend is observed: the coincidence count between detector bars decreases as shielding increases. Panel 1 (blue circles) and Panel 2 (red squares) exhibit consistent behaviour within their statistical uncertainties. The smooth decrease confirms the effective suppression of low-energy background particles. The green diamond markers correspond to the results obtained from simulations, which show a comparable attenuation trend.}
    \label{fig:pb_results}
\end{figure}
Figure~\ref{fig:pb_results} presents the normalised flux measured by both detection panels as a function of lead shielding thickness. As the lead thickness increases, the coincidence count between scintillator bars steadily decreases, showing a clear attenuation trend. Panels 1 (blue circles) and 2 (red squares) exhibit consistent behaviour within statistical uncertainties. The data follows a smooth, monotonic decline in flux, confirming the expected suppression of low-energy (soft) components. Simulations reproduce this qualitative behaviour, which validates both our experimental methodology and the detector model implemented in MEIGA. This agreement reinforces the reliability of our setup for studying shielding effects and characterising detector performance under different environmental conditions.

Simulated results show slightly more substantial attenuation than the experimental measurements. This difference likely stems from the structural environment where we conducted the experiments. We performed the measurements on the third floor of a reinforced concrete building, which naturally absorbs a portion of the soft component of cosmic rays before they reach the detector. In contrast, the simulation assumes an unshielded, surface-level cosmic ray spectrum impinging directly on the detector. Consequently, lead shielding in the simulation removes a more significant fraction of soft particles than in the actual setup, where the building structure has already filtered much of that component. Despite this discrepancy, the experimental and simulated data show the same overall attenuation trend with increasing lead thickness. This consistency confirms the system's sensitivity to shielding materials and supports the accuracy of our detector model.

\begin{figure}
    \centering
    \includegraphics[width=0.35\textwidth]{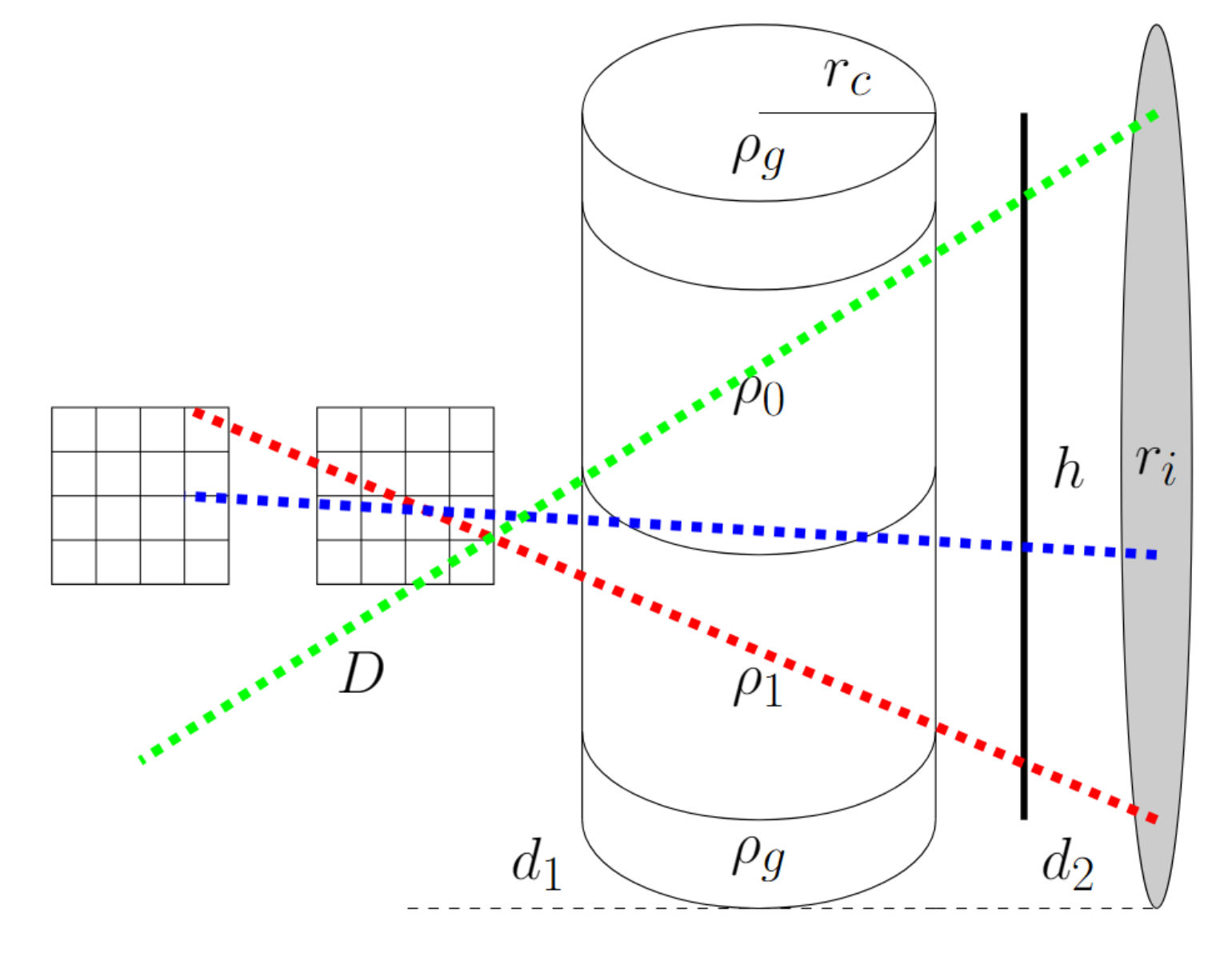}
    \includegraphics[width=0.35\textwidth]{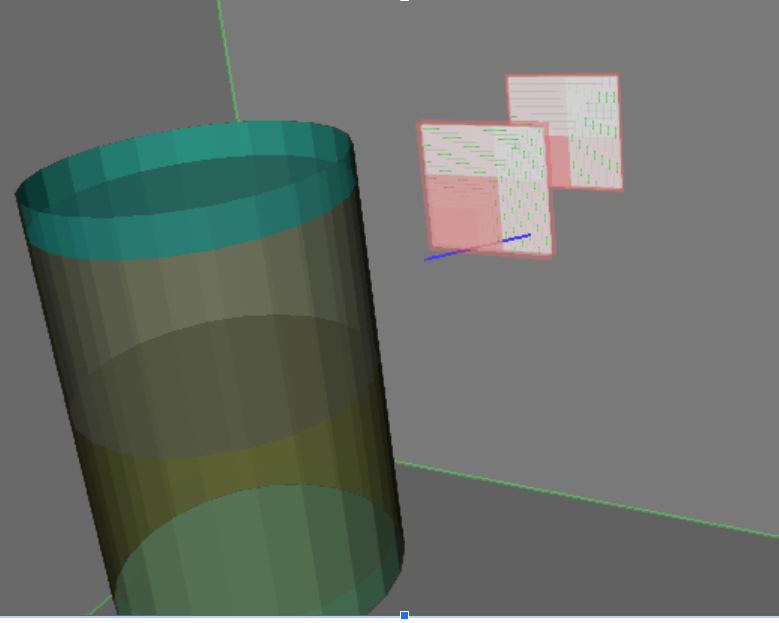}
    \caption{(Up) Schematic representation of the muography simulation setup. The diagram shows the injection ray, the hydrotreatment tower with its two density interfaces, the hodoscope detector system, and several simulated muon trajectories intersecting the tower and reaching the detector. (Down) Visual rendering of the hydrotreatment tower model implemented in the MEIGA framework.}
    \label{fig:tower_model}
\end{figure}

\section{Hydrotreatment Tower}
Hydrotreatment towers play a vital role in the petroleum and petrochemical industries, particularly in purifying and upgrading fuels in general. These reactors operate under high-pressure and high-temperature conditions to drive catalytic reactions that remove impurities like sulfur, nitrogen, and heavy metals. This section presents a detailed model of a hydrotreatment tower as implemented in our muography simulation framework. The primary objective is to evaluate the feasibility of muon imaging for resolving internal structural features—especially variations in catalyst density—that critically impact reactor performance.

We modelled the hydrotreatment tower as a cylindrical structure divided into multiple sections, each corresponding to different materials and functional zones. At the core lies the catalyst bed, surrounded by structural support and thermal insulation layers. The catalyst is represented by a typical $\text{NiMo/Al}_2\text{O}_3$ formulation, composed of $40.22$\% $\text{Al}$, 43.3\% $\text{O}$, $13.34$\% $\text{Mo}$, and $3.14$\% $\text{Ni}$, accurately reflecting the chemical composition used in industrial hydroprocessing applications.

The model is $144$~cm tall with a fixed radius of $50$~cm. It includes an initial inert section ($12$~cm), followed by two catalyst beds of equal length ($60$~cm each) but differing in density: the first has a standard density of $1.2$~g/cm$^3$, and the second a higher density of $3.1$~g/cm$^3$. A final $12$~cm inert section completes the structure. This configuration mimics the internal layout of a functioning hydrotreatment tower and allows us to simulate how muons interact with different density regions inside the reactor.

\begin{figure*}
    \centering
    \includegraphics[width=0.32\textwidth]{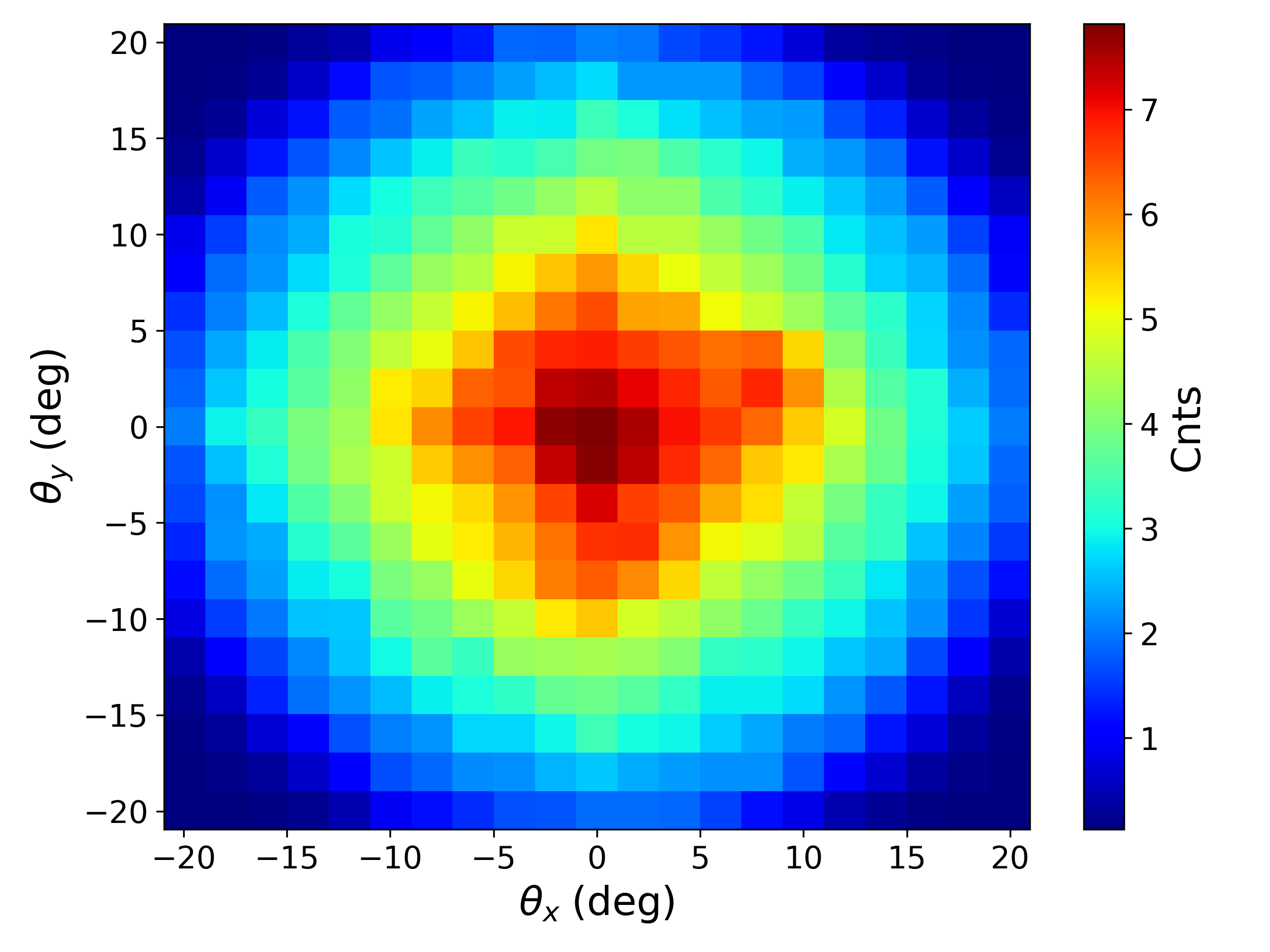}
    \includegraphics[width=0.32\textwidth]{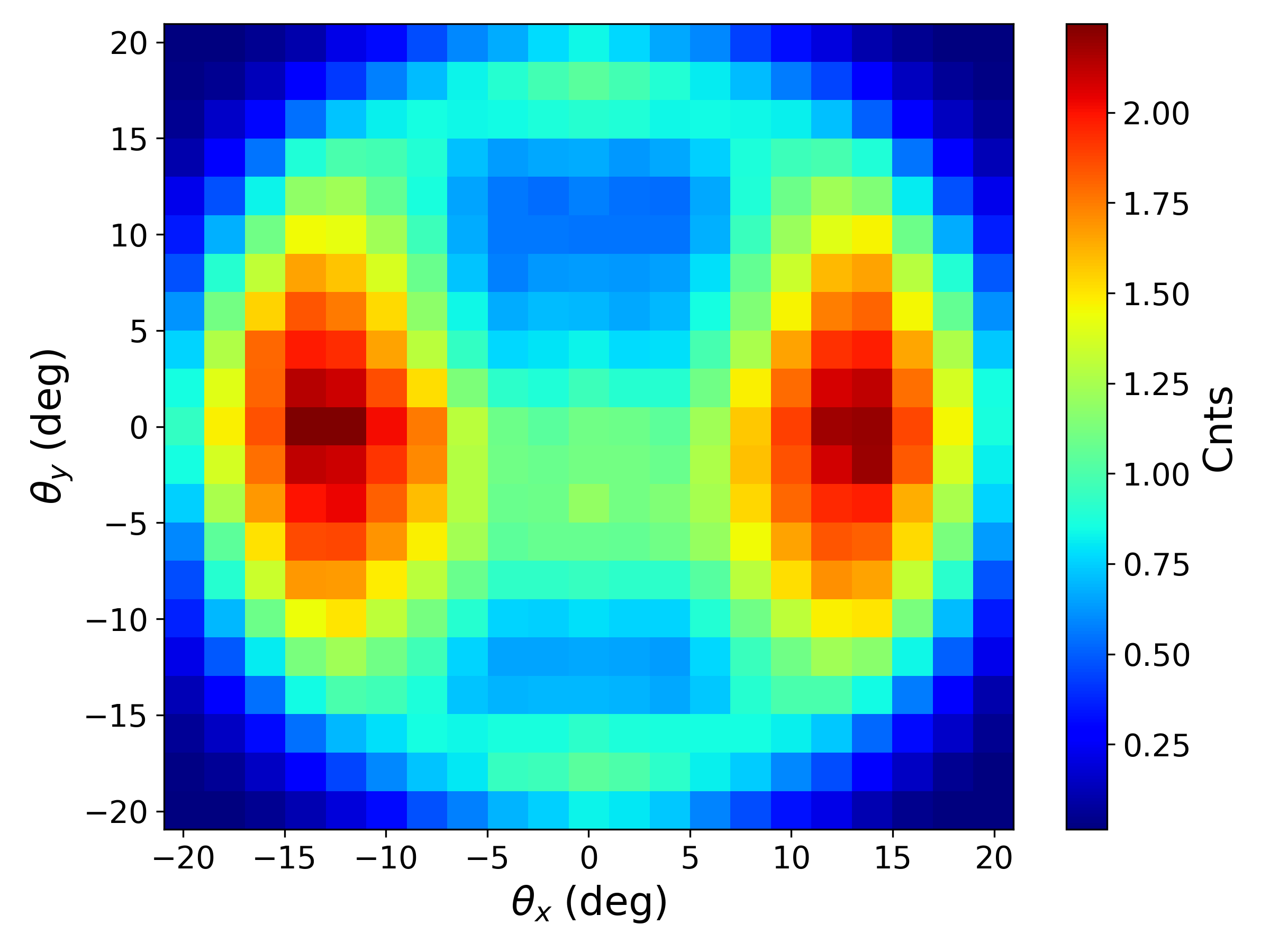}
    \includegraphics[width=0.35\textwidth]{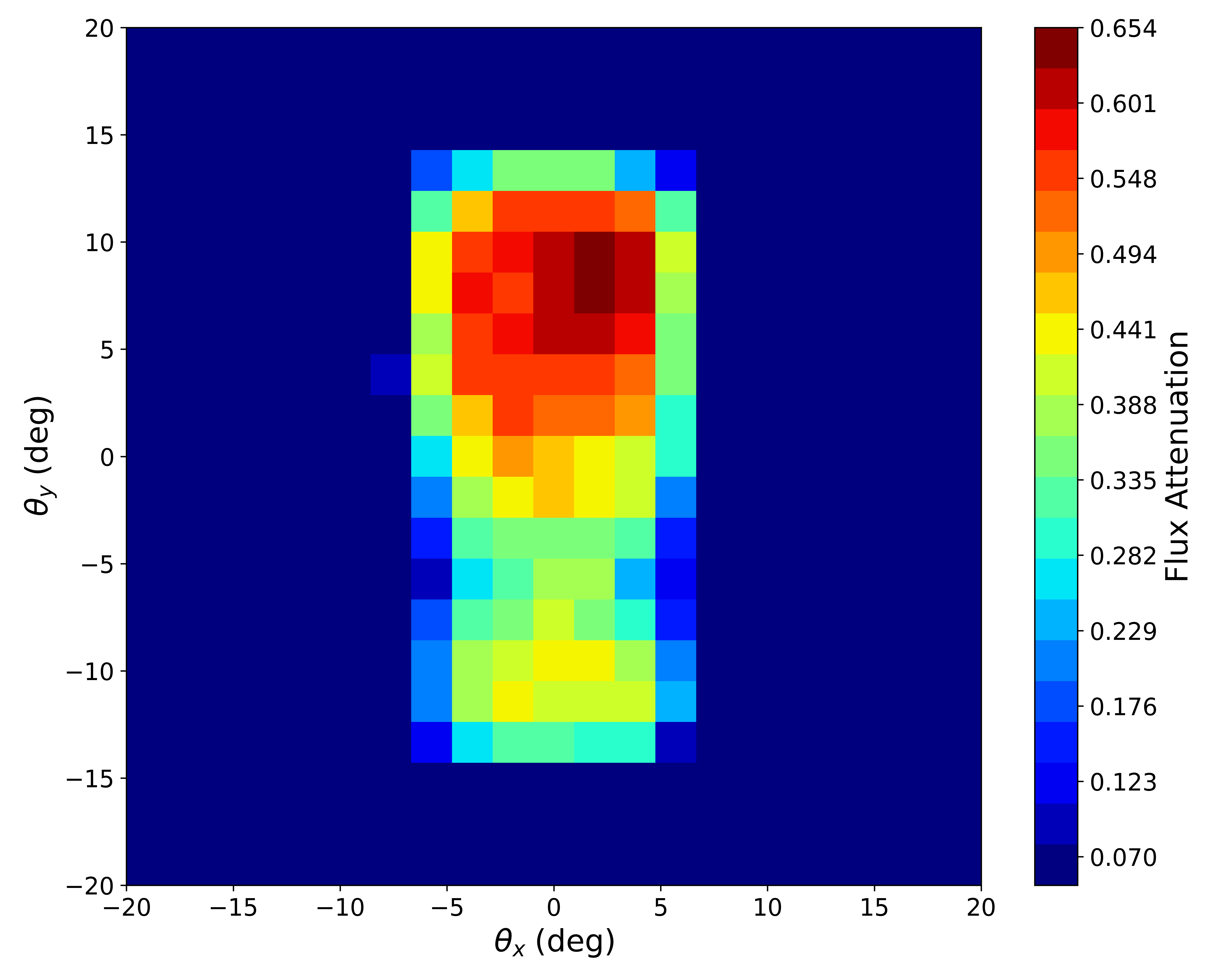}
    \caption{(Left) Angular flux distribution from cosmic rays in the absence of the tower (open sky simulation). (Center) Angular distribution of muons after traversing the hydrotreatment tower model. (Right) Relative difference map computed as $(F_\text{open} - F_\text{tower})/F_\text{open}$, where $F_\text{open}$ and $F_\text{tower}$ correspond to the angular fluxes without and with the tower, respectively. The final map reveals the attenuation signature induced by the catalyst density contrast.}
    \label{fig:tower_flux_maps}
\end{figure*}

To evaluate the potential of muography for imaging internal structures in a hydrotreatment units, we specifically designed the catalyst bed to feature these two distinct density regions. This contrast enables the study of how spatial variations in material density affect muon attenuation and detection efficiency. By analysing the differential attenuation of muons across these regions, we can assess the technique’s ability to resolve internal features critical to reactor performance, such as catalyst bed compaction, degradation, or uneven distribution.

The reactor includes several structural and thermal protection layers: a $0.2$~cm-thick steel shell for mechanical strength, a $0.3$~cm-thick insulating calorifuge, and a $0.1$~cm-thick aluminium cladding. These layers replicate the mechanical and thermal constraints encountered in real industrial settings, where reactors operate under high pressures and elevated temperatures.

To represent the fluid contents within the reactor, we incorporated common hydrocarbons into the simulation, including paraffin $\text{C}_n\text{H}_{2n+2}$, benzene $\text{C}_6\text{H}_6$, and diesel, the latter modelled as a mixture of 75\% paraffin and 25\% benzene—consistent with typical refining compositions. Paraffin and benzene are assigned elemental compositions of 84.7\% C / 15.3\% H and 92.3\% C / 7.7\% H, respectively. In addition, we included trace elements such as S, Ca, and Na to mimic impurities and secondary components often found in industrial environments. Structural support materials, including stainless steel (Fe, Ni, Cr) and aluminium (Al), complete the tower's framework, providing the necessary thermal resistance and mechanical stability.

We configured the simulation to inject muons from a circular flux source with a radius of $1.8$~m, ensuring a wide range of incident angles for comprehensive coverage. Muons enter the system $5$~cm from the tower's outer surface, allowing precise tracking of their interactions with internal components. The towering center is positioned $100$~cm in front of the first detection panel, and the second panel is placed $100$~cm behind the first, creating a total $100$~cm baseline for muon trajectory reconstruction.

To quantify muon attenuation, we conducted two simulation runs. First, we reconstructed the open-sky angular flux distribution using only the hodoscope geometry (Figure~\ref{fig:tower_flux_maps}, left). By analysing coincident hits between scintillator bars, we identified the incoming direction of each muon.  In the second run, we introduced the full hydrotreatment tower model into the simulation and repeated the angular flux reconstruction (Figure~\ref{fig:tower_flux_maps}, centre). The comparison between the two flux maps clearly shows the attenuation caused by the internal structures of the tower. We then computed a relative attenuation map by subtracting the tower-inclusive flux from the open-sky flux and normalising it to the open-sky case. The resulting image sketched in Figure~\ref{fig:tower_flux_maps}, (right) reveals angular sectors with significant attenuation, directly corresponding to the denser catalyst zones inside the reactor. These results demonstrate that muography can resolve density contrasts within complex industrial systems, supporting its application for non-invasive internal diagnostics.

We performed these simulations assuming a vertical muon flux with an equivalent exposure time of $20$~hours. However, detector alignment with the vertical axis is not always achievable in practical deployments. When the hodoscope is tilted at $60^o$ from vertical, the effective muon acceptance decreases, requiring approximately $36$~hours to achieve comparable image quality. For a more horizontal configuration, such as $30^o$ from horizontal, the required exposure time increases significantly—up to $8$~days. Despite these extended acquisition periods, the timeframe remains practical for routine industrial monitoring, with timescales ranging from days to weeks. These findings highlight the importance of detector orientation in optimizing acquisition time and image resolution, especially in constrained environments like oil refineries.
\section{Conclusions and Feasibility Study}
This study successfully validated a scintillator-based hodoscope for muon detection, demonstrating its suitability for muography applications in industrial environments such as oil refineries. Through a comprehensive calibration procedure—based on the staircase method—we ensured uniform response and effective noise suppression across 60 independent channels. The strong agreement between experimental data and simulations confirmed the detector's stability, reliability, and precision in real-world conditions.

Open-sky measurements showed stable muon flux detection, with values of approximately $167 \pm 16$ particles/m$^2$/s for Panel 1 and $179 \pm 17$ particles/m$^2$/s for Panel 2.  The absence of correlations with environmental light or temperature variations affirmed the system's robustness against external noise sources. Coincidence maps revealed angular distributions consistent with the system's geometric acceptance, validating the angular reconstruction capabilities crucial for muographic imaging.

Lead shielding tests further reinforced the system's performance, demonstrating apparent flux attenuation with increasing material thickness. The experimental trends closely matched simulated results, confirming the detector's sensitivity to soft-component suppression. These results validated the MEIGA simulation framework, especially when coupled with realistic secondary flux models generated using ARTI.

We modelled a hydrotreatment tower using the validated simulation framework to explore the practical utility of muography in the oil industry. The model incorporated realistic structural materials and hydroprocessing catalysts with distinct density zones. Simulated muon flux maps revealed angular attenuation patterns corresponding to internal heterogeneities within the tower. These findings confirm muography's capability to resolve density contrasts and identify structural anomalies, such as catalyst bed compaction or uneven flow paths.

We also evaluated the influence of detector orientation on imaging efficiency. A vertical configuration allowed high-quality imaging within 20 hours of exposure, whereas $60^o$ and $30^o$ inclinations extended the required time to approximately $36$~hours and $8$~days, respectively. These results highlight the importance of maximising vertical muon flux acceptance and optimising detector geometry to balance resolution and acquisition time—key factors for real-time or routine monitoring in industrial applications.

Beyond this proof-of-concept, the study demonstrates that muography is a feasible, non-invasive technique for structural diagnostics in the oil and gas sector. It is particularly well-suited for monitoring hydrotreatment reactors, where internal conditions such as catalyst bed settling, fouling, or material degradation can significantly impact process efficiency. With exposure times ranging from hours to a few days, muon imaging offers a cost-effective complement or alternative to traditional inspection methods, especially in impractical limited access or shutdowns.

Future work will focus on integrating real-time muographic monitoring systems into operational facilities, enhancing reconstruction algorithms for higher-resolution imaging, and expanding simulation tools to incorporate time-evolving internal structures. By bridging high-energy physics and industrial process monitoring, muography holds significant promise as a transformative tool for next-generation infrastructure diagnostics in energy-intensive sectors.
\section{Acknowledgement}
The authors also grateful for the funding of this work in the framework of the project ``Integración de muongrafía con métodos geofísicos estándar para la construcción de un modelo 3-D de densidad: aplicación al Volcán Cerro Machín.'', under project 82242 of the call 890 of 2020, Minciencias, administered through the ICETEX contract 2022-0718. R.A.M.R. gratefully thanks the internship program of the ERASMUS+ project, Latin-American alliance for Capacity buildiNG in Advance physics (LA-CoNGA physics), where this paper’s first ideas and calculations began. We also acknowledge the
computational support from the Universidad Industrial de Santander (SC3UIS) High Performance and Scientific Computing Centre.

\bibliography{referencias}

\begin{thebibliography}{26}%
\makeatletter
\providecommand \@ifxundefined [1]{%
 \@ifx{#1\undefined}
}%
\providecommand \@ifnum [1]{%
 \ifnum #1\expandafter \@firstoftwo
 \else \expandafter \@secondoftwo
 \fi
}%
\providecommand \@ifx [1]{%
 \ifx #1\expandafter \@firstoftwo
 \else \expandafter \@secondoftwo
 \fi
}%
\providecommand \natexlab [1]{#1}%
\providecommand \enquote  [1]{``#1''}%
\providecommand \bibnamefont  [1]{#1}%
\providecommand \bibfnamefont [1]{#1}%
\providecommand \citenamefont [1]{#1}%
\providecommand \href@noop [0]{\@secondoftwo}%
\providecommand \href [0]{\begingroup \@sanitize@url \@href}%
\providecommand \@href[1]{\@@startlink{#1}\@@href}%
\providecommand \@@href[1]{\endgroup#1\@@endlink}%
\providecommand \@sanitize@url [0]{\catcode `\\12\catcode `\$12\catcode `\&12\catcode `\#12\catcode `\^12\catcode `\_12\catcode `\%12\relax}%
\providecommand \@@startlink[1]{}%
\providecommand \@@endlink[0]{}%
\providecommand \url  [0]{\begingroup\@sanitize@url \@url }%
\providecommand \@url [1]{\endgroup\@href {#1}{\urlprefix }}%
\providecommand \urlprefix  [0]{URL }%
\providecommand \Eprint [0]{\href }%
\providecommand \doibase [0]{http://dx.doi.org/}%
\providecommand \selectlanguage [0]{\@gobble}%
\providecommand \bibinfo  [0]{\@secondoftwo}%
\providecommand \bibfield  [0]{\@secondoftwo}%
\providecommand \translation [1]{[#1]}%
\providecommand \BibitemOpen [0]{}%
\providecommand \bibitemStop [0]{}%
\providecommand \bibitemNoStop [0]{.\EOS\space}%
\providecommand \EOS [0]{\spacefactor3000\relax}%
\providecommand \BibitemShut  [1]{\csname bibitem#1\endcsname}%
\let\auto@bib@innerbib\@empty
\bibitem [{\citenamefont {Gaisser}, \citenamefont {Engel},\ and\ \citenamefont {Resconi}(2016)}]{GaisserEngelResconi2016}%
  \BibitemOpen
  \bibfield  {author} {\bibinfo {author} {\bibfnamefont {T.}~\bibnamefont {Gaisser}}, \bibinfo {author} {\bibfnamefont {R.}~\bibnamefont {Engel}}, \ and\ \bibinfo {author} {\bibfnamefont {E.}~\bibnamefont {Resconi}},\ }\href {\doibase 10.1017/CBO9781139192194} {\emph {\bibinfo {title} {Cosmic Rays and Particle Physics}}}\ (\bibinfo  {publisher} {Cambridge University Press},\ \bibinfo {year} {2016})\BibitemShut {NoStop}%
\bibitem [{\citenamefont {Tanaka}\ \emph {et~al.}(2023)\citenamefont {Tanaka}, \citenamefont {Bozza}, \citenamefont {Bross}, \citenamefont {Cantoni}, \citenamefont {Catalano}, \citenamefont {Cerretto}, \citenamefont {Giammanco}, \citenamefont {Gluyas}, \citenamefont {Gnesi}, \citenamefont {Holma}, \citenamefont {Kin}, \citenamefont {Roche}, \citenamefont {Leone}, \citenamefont {Liu}, \citenamefont {Presti}, \citenamefont {Marteau}, \citenamefont {Matsushima}, \citenamefont {Oláh}, \citenamefont {Polukhina}, \citenamefont {Ramakrishna}, \citenamefont {Sellone}, \citenamefont {Shinohara}, \citenamefont {Steigerwald}, \citenamefont {Sumiya}, \citenamefont {Thompson}, \citenamefont {Tioukov}, \citenamefont {Yokota},\ and\ \citenamefont {Varga}}]{TanakaEtal2023}%
  \BibitemOpen
  \bibfield  {author} {\bibinfo {author} {\bibfnamefont {H.}~\bibnamefont {Tanaka}}, \bibinfo {author} {\bibfnamefont {C.}~\bibnamefont {Bozza}}, \bibinfo {author} {\bibfnamefont {A.}~\bibnamefont {Bross}}, \bibinfo {author} {\bibfnamefont {E.}~\bibnamefont {Cantoni}}, \bibinfo {author} {\bibfnamefont {O.}~\bibnamefont {Catalano}}, \bibinfo {author} {\bibfnamefont {G.}~\bibnamefont {Cerretto}}, \bibinfo {author} {\bibfnamefont {A.}~\bibnamefont {Giammanco}}, \bibinfo {author} {\bibfnamefont {J.}~\bibnamefont {Gluyas}}, \bibinfo {author} {\bibfnamefont {I.}~\bibnamefont {Gnesi}}, \bibinfo {author} {\bibfnamefont {M.}~\bibnamefont {Holma}}, \bibinfo {author} {\bibfnamefont {T.}~\bibnamefont {Kin}}, \bibinfo {author} {\bibfnamefont {I.}~\bibnamefont {Roche}}, \bibinfo {author} {\bibfnamefont {G.}~\bibnamefont {Leone}}, \bibinfo {author} {\bibfnamefont {Z.}~\bibnamefont {Liu}}, \bibinfo {author} {\bibfnamefont {D.~L.}\ \bibnamefont {Presti}}, \bibinfo {author} {\bibfnamefont {J.}~\bibnamefont {Marteau}}, \bibinfo
  {author} {\bibfnamefont {J.}~\bibnamefont {Matsushima}}, \bibinfo {author} {\bibfnamefont {L.}~\bibnamefont {Oláh}}, \bibinfo {author} {\bibfnamefont {N.}~\bibnamefont {Polukhina}}, \bibinfo {author} {\bibfnamefont {S.~S.}\ \bibnamefont {Ramakrishna}}, \bibinfo {author} {\bibfnamefont {M.}~\bibnamefont {Sellone}}, \bibinfo {author} {\bibfnamefont {A.}~\bibnamefont {Shinohara}}, \bibinfo {author} {\bibfnamefont {S.}~\bibnamefont {Steigerwald}}, \bibinfo {author} {\bibfnamefont {K.}~\bibnamefont {Sumiya}}, \bibinfo {author} {\bibfnamefont {L.}~\bibnamefont {Thompson}}, \bibinfo {author} {\bibfnamefont {V.}~\bibnamefont {Tioukov}}, \bibinfo {author} {\bibfnamefont {Y.}~\bibnamefont {Yokota}}, \ and\ \bibinfo {author} {\bibfnamefont {D.}~\bibnamefont {Varga}},\ }\bibfield  {title} {\enquote {\bibinfo {title} {Muography},}\ }\href@noop {} {\bibfield  {journal} {\bibinfo  {journal} {Nature Reviews Methods Primers}\ }\textbf {\bibinfo {volume} {3}},\ \bibinfo {pages} {88} (\bibinfo {year} {2023})}\BibitemShut
  {NoStop}%
\bibitem [{\citenamefont {Tanaka}\ \emph {et~al.}(2007)\citenamefont {Tanaka}, \citenamefont {Nakano}, \citenamefont {Takahashi}, \citenamefont {Yoshida}, \citenamefont {Ohshima}, \citenamefont {Maekawa}, \citenamefont {Watanabe},\ and\ \citenamefont {Niwa}}]{TanakaEtal2007b}%
  \BibitemOpen
  \bibfield  {author} {\bibinfo {author} {\bibfnamefont {H.}~\bibnamefont {Tanaka}}, \bibinfo {author} {\bibfnamefont {T.}~\bibnamefont {Nakano}}, \bibinfo {author} {\bibfnamefont {S.}~\bibnamefont {Takahashi}}, \bibinfo {author} {\bibfnamefont {J.}~\bibnamefont {Yoshida}}, \bibinfo {author} {\bibfnamefont {H.}~\bibnamefont {Ohshima}}, \bibinfo {author} {\bibfnamefont {T.}~\bibnamefont {Maekawa}}, \bibinfo {author} {\bibfnamefont {H.}~\bibnamefont {Watanabe}}, \ and\ \bibinfo {author} {\bibfnamefont {K.}~\bibnamefont {Niwa}},\ }\bibfield  {title} {\enquote {\bibinfo {title} {Imaging the conduit size of the dome with cosmic-ray muons: The structure beneath showa-shinzan lava dome, japan},}\ }\href@noop {} {\bibfield  {journal} {\bibinfo  {journal} {Geophysical Research Letters}\ }\textbf {\bibinfo {volume} {34}} (\bibinfo {year} {2007})}\BibitemShut {NoStop}%
\bibitem [{\citenamefont {Pezzotti}\ \emph {et~al.}(2025)\citenamefont {Pezzotti}, \citenamefont {Cifarelli}, \citenamefont {Corradetti}, \citenamefont {Orlandi},\ and\ \citenamefont {Sedoski}}]{PezzottiEtal2025}%
  \BibitemOpen
  \bibfield  {author} {\bibinfo {author} {\bibfnamefont {L.}~\bibnamefont {Pezzotti}}, \bibinfo {author} {\bibfnamefont {D.}~\bibnamefont {Cifarelli}}, \bibinfo {author} {\bibfnamefont {D.}~\bibnamefont {Corradetti}}, \bibinfo {author} {\bibfnamefont {A.}~\bibnamefont {Orlandi}}, \ and\ \bibinfo {author} {\bibfnamefont {A.}~\bibnamefont {Sedoski}},\ }\bibfield  {title} {\enquote {\bibinfo {title} {A new method for structural diagnostics with muon tomography and deep learning},}\ }\href {https://arxiv.org/abs/2502.03339} {\bibfield  {journal} {\bibinfo  {journal} {arXiv preprint}\ } (\bibinfo {year} {2025})},\ \Eprint {http://arxiv.org/abs/2502.03339} {2502.03339} \BibitemShut {NoStop}%
\bibitem [{\citenamefont {Pe{\~{n}}a-Rodr{\'{\i}}guez}\ \emph {et~al.}(2020)\citenamefont {Pe{\~{n}}a-Rodr{\'{\i}}guez}, \citenamefont {Pisco-Guabave}, \citenamefont {Sierra-Porta}, \citenamefont {Su{\'{a}}rez-Dur{\'{a}}n}, \citenamefont {Arenas-Fl{\'{o}}rez}, \citenamefont {P{\'{e}}rez-Archila}, \citenamefont {Sanabria-G{\'{o}}mez}, \citenamefont {Asorey},\ and\ \citenamefont {N{\'{u}}{\~{n}}ez}}]{PenarodriguezEtal2020}%
  \BibitemOpen
  \bibfield  {author} {\bibinfo {author} {\bibfnamefont {J.}~\bibnamefont {Pe{\~{n}}a-Rodr{\'{\i}}guez}}, \bibinfo {author} {\bibfnamefont {J.}~\bibnamefont {Pisco-Guabave}}, \bibinfo {author} {\bibfnamefont {D.}~\bibnamefont {Sierra-Porta}}, \bibinfo {author} {\bibfnamefont {M.}~\bibnamefont {Su{\'{a}}rez-Dur{\'{a}}n}}, \bibinfo {author} {\bibfnamefont {M.}~\bibnamefont {Arenas-Fl{\'{o}}rez}}, \bibinfo {author} {\bibfnamefont {L.}~\bibnamefont {P{\'{e}}rez-Archila}}, \bibinfo {author} {\bibfnamefont {J.}~\bibnamefont {Sanabria-G{\'{o}}mez}}, \bibinfo {author} {\bibfnamefont {H.}~\bibnamefont {Asorey}}, \ and\ \bibinfo {author} {\bibfnamefont {L.}~\bibnamefont {N{\'{u}}{\~{n}}ez}},\ }\bibfield  {title} {\enquote {\bibinfo {title} {Design and construction of mute: a hybrid muon telescope to study colombian volcanoes},}\ }\href@noop {} {\bibfield  {journal} {\bibinfo  {journal} {Journal of Instrumentation}\ }\textbf {\bibinfo {volume} {15}},\ \bibinfo {pages} {P09006--P09006} (\bibinfo {year}
  {2020})}\BibitemShut {NoStop}%
\bibitem [{\citenamefont {Calder{\'o}n-Ardila}\ \emph {et~al.}(2020)\citenamefont {Calder{\'o}n-Ardila}, \citenamefont {Almela}, \citenamefont {G{\'o}mez-Berisso}, \citenamefont {Sedoski}, \citenamefont {Varela}, \citenamefont {Vesga-Ram{\'\i}rez},\ and\ \citenamefont {Asorey}}]{CalderonEtal2020}%
  \BibitemOpen
  \bibfield  {author} {\bibinfo {author} {\bibfnamefont {R.}~\bibnamefont {Calder{\'o}n-Ardila}}, \bibinfo {author} {\bibfnamefont {A.}~\bibnamefont {Almela}}, \bibinfo {author} {\bibfnamefont {M.}~\bibnamefont {G{\'o}mez-Berisso}}, \bibinfo {author} {\bibfnamefont {A.}~\bibnamefont {Sedoski}}, \bibinfo {author} {\bibfnamefont {C.}~\bibnamefont {Varela}}, \bibinfo {author} {\bibfnamefont {A.}~\bibnamefont {Vesga-Ram{\'\i}rez}}, \ and\ \bibinfo {author} {\bibfnamefont {H.}~\bibnamefont {Asorey}},\ }\bibfield  {title} {\enquote {\bibinfo {title} {Study of spatial resolution of muon hodoscopes for muography applications in geophysics},}\ }\href@noop {} {\bibfield  {journal} {\bibinfo  {journal} {arXiv preprint arXiv:2006.03165}\ } (\bibinfo {year} {2020})}\BibitemShut {NoStop}%
\bibitem [{\citenamefont {Tishevsky}\ \emph {et~al.}(2024{\natexlab{a}})\citenamefont {Tishevsky}, \citenamefont {Dubinin}, \citenamefont {Isupov}, \citenamefont {Ladygin}, \citenamefont {Nigmatkulov}, \citenamefont {Reznikov}, \citenamefont {Teterin}, \citenamefont {Volkov},\ and\ \citenamefont {Zakharov}}]{Tishevsky2024_SPD}%
  \BibitemOpen
  \bibfield  {author} {\bibinfo {author} {\bibfnamefont {A.~V.}\ \bibnamefont {Tishevsky}}, \bibinfo {author} {\bibfnamefont {F.~A.}\ \bibnamefont {Dubinin}}, \bibinfo {author} {\bibfnamefont {A.~Y.}\ \bibnamefont {Isupov}}, \bibinfo {author} {\bibfnamefont {V.~P.}\ \bibnamefont {Ladygin}}, \bibinfo {author} {\bibfnamefont {G.~A.}\ \bibnamefont {Nigmatkulov}}, \bibinfo {author} {\bibfnamefont {S.~G.}\ \bibnamefont {Reznikov}}, \bibinfo {author} {\bibfnamefont {P.~E.}\ \bibnamefont {Teterin}}, \bibinfo {author} {\bibfnamefont {I.~S.}\ \bibnamefont {Volkov}}, \ and\ \bibinfo {author} {\bibfnamefont {A.~M.}\ \bibnamefont {Zakharov}},\ }\bibfield  {title} {\enquote {\bibinfo {title} {The spd beam-beam counter scintillation detector prototype tests with fers-5200 front-end readout system},}\ }\href {\doibase 10.1134/S1547477124701188} {\bibfield  {journal} {\bibinfo  {journal} {Physics of Particles and Nuclei Letters}\ }\textbf {\bibinfo {volume} {21}},\ \bibinfo {pages} {723–726} (\bibinfo {year}
  {2024}{\natexlab{a}})}\BibitemShut {NoStop}%
\bibitem [{\citenamefont {Sarmiento-Cano}\ \emph {et~al.}(2022)\citenamefont {Sarmiento-Cano}, \citenamefont {Suárez-Durán}, \citenamefont {Calderón-Ardila}, \citenamefont {Vásquez-Ramírez}, \citenamefont {Jaimes-Motta}, \citenamefont {Núñez}, \citenamefont {Dasso}, \citenamefont {Sidelnik},\ and\ \citenamefont {Asorey}}]{SarmientoCano_2022}%
  \BibitemOpen
  \bibfield  {author} {\bibinfo {author} {\bibfnamefont {C.}~\bibnamefont {Sarmiento-Cano}}, \bibinfo {author} {\bibfnamefont {M.}~\bibnamefont {Suárez-Durán}}, \bibinfo {author} {\bibfnamefont {R.}~\bibnamefont {Calderón-Ardila}}, \bibinfo {author} {\bibfnamefont {A.}~\bibnamefont {Vásquez-Ramírez}}, \bibinfo {author} {\bibfnamefont {A.}~\bibnamefont {Jaimes-Motta}}, \bibinfo {author} {\bibfnamefont {L.~A.}\ \bibnamefont {Núñez}}, \bibinfo {author} {\bibfnamefont {S.}~\bibnamefont {Dasso}}, \bibinfo {author} {\bibfnamefont {I.}~\bibnamefont {Sidelnik}}, \ and\ \bibinfo {author} {\bibfnamefont {H.}~\bibnamefont {Asorey}},\ }\bibfield  {title} {\enquote {\bibinfo {title} {The arti framework: cosmic rays atmospheric background simulations},}\ }\href {\doibase 10.1140/epjc/s10052-022-10883-z} {\bibfield  {journal} {\bibinfo  {journal} {European Physical Journal C}\ } (\bibinfo {year} {2022}),\ 10.1140/epjc/s10052-022-10883-z}\BibitemShut {NoStop}%
\bibitem [{\citenamefont {Taboada}\ \emph {et~al.}(2022)\citenamefont {Taboada}, \citenamefont {Sarmiento-Cano}, \citenamefont {Sedoski},\ and\ \citenamefont {Asorey}}]{TaboadaEtal2022}%
  \BibitemOpen
  \bibfield  {author} {\bibinfo {author} {\bibfnamefont {A.}~\bibnamefont {Taboada}}, \bibinfo {author} {\bibfnamefont {C.}~\bibnamefont {Sarmiento-Cano}}, \bibinfo {author} {\bibfnamefont {A.}~\bibnamefont {Sedoski}}, \ and\ \bibinfo {author} {\bibfnamefont {H.}~\bibnamefont {Asorey}},\ }\bibfield  {title} {\enquote {\bibinfo {title} {Meiga, a dedicated framework used for muography applications},}\ }\href {\doibase https://doi.org/10.31526/jais.2022.266} {\bibfield  {journal} {\bibinfo  {journal} {Journal of Advanced Instrumentation in Science}\ }\textbf {\bibinfo {volume} {2022}} (\bibinfo {year} {2022}),\ https://doi.org/10.31526/jais.2022.266}\BibitemShut {NoStop}%
\bibitem [{\citenamefont {Pe{\~n}a-Rodr{\'\i}guez}\ \emph {et~al.}(2024)\citenamefont {Pe{\~n}a-Rodr{\'\i}guez}, \citenamefont {Jaimes-Teher{\'a}n}, \citenamefont {Dlaikan-Castillo},\ and\ \citenamefont {N{\'u}{\~n}ez}}]{PenaRodriguezEtal2024}%
  \BibitemOpen
  \bibfield  {author} {\bibinfo {author} {\bibfnamefont {J.}~\bibnamefont {Pe{\~n}a-Rodr{\'\i}guez}}, \bibinfo {author} {\bibfnamefont {J.}~\bibnamefont {Jaimes-Teher{\'a}n}}, \bibinfo {author} {\bibfnamefont {K.}~\bibnamefont {Dlaikan-Castillo}}, \ and\ \bibinfo {author} {\bibfnamefont {L.}~\bibnamefont {N{\'u}{\~n}ez}},\ }\bibfield  {title} {\enquote {\bibinfo {title} {Muysc: an end-to-end muography simulation toolbox},}\ }\href@noop {} {\bibfield  {journal} {\bibinfo  {journal} {Geophysical Journal International}\ }\textbf {\bibinfo {volume} {237}},\ \bibinfo {pages} {540--556} (\bibinfo {year} {2024})}\BibitemShut {NoStop}%
\bibitem [{\citenamefont {Furimsky}\ and\ \citenamefont {Massoth}(1999)}]{Furimsky_1999}%
  \BibitemOpen
  \bibfield  {author} {\bibinfo {author} {\bibfnamefont {E.}~\bibnamefont {Furimsky}}\ and\ \bibinfo {author} {\bibfnamefont {F.~E.}\ \bibnamefont {Massoth}},\ }\bibfield  {title} {\enquote {\bibinfo {title} {Deactivation of hydroprocessing catalyst},}\ }\href {\doibase 10.1016/s0920-5861(99)00096-6} {\bibfield  {journal} {\bibinfo  {journal} {Catalysis Today}\ } (\bibinfo {year} {1999}),\ 10.1016/s0920-5861(99)00096-6}\BibitemShut {NoStop}%
\bibitem [{\citenamefont {Argyle}\ and\ \citenamefont {Bartholomew}(2015)}]{Argyle_2015}%
  \BibitemOpen
  \bibfield  {author} {\bibinfo {author} {\bibfnamefont {M.~D.}\ \bibnamefont {Argyle}}\ and\ \bibinfo {author} {\bibfnamefont {C.~H.}\ \bibnamefont {Bartholomew}},\ }\bibfield  {title} {\enquote {\bibinfo {title} {Heterogeneous catalyst deactivation and regeneration: A review},}\ }\href {\doibase 10.3390/catal5010145} {\bibfield  {journal} {\bibinfo  {journal} {Catalysts}\ } (\bibinfo {year} {2015}),\ 10.3390/catal5010145}\BibitemShut {NoStop}%
\bibitem [{\citenamefont {Duarte}, \citenamefont {Garzón},\ and\ \citenamefont {Baldovino-Medrano}(2019)}]{Duarte_2019}%
  \BibitemOpen
  \bibfield  {author} {\bibinfo {author} {\bibfnamefont {L.~J.}\ \bibnamefont {Duarte}}, \bibinfo {author} {\bibfnamefont {L.}~\bibnamefont {Garzón}}, \ and\ \bibinfo {author} {\bibfnamefont {V.~G.}\ \bibnamefont {Baldovino-Medrano}},\ }\bibfield  {title} {\enquote {\bibinfo {title} {An analysis of the physicochemical properties of spent catalysts from an industrial hydrotreating unit},}\ }\href {\doibase 10.1016/j.cattod.2019.05.025} {\bibfield  {journal} {\bibinfo  {journal} {Catalysis Today}\ } (\bibinfo {year} {2019}),\ 10.1016/j.cattod.2019.05.025}\BibitemShut {NoStop}%
\bibitem [{Note1()}]{Note1}%
  \BibitemOpen
  \bibinfo {note} {\protect \url {https://github.com/lagoproject/arti}}\BibitemShut {NoStop}%
\bibitem [{\citenamefont {Asorey}\ and\ \citenamefont {et~al.}(2018)}]{AsoreyNunezSuarez2018}%
  \BibitemOpen
  \bibfield  {author} {\bibinfo {author} {\bibfnamefont {H.}~\bibnamefont {Asorey}}\ and\ \bibinfo {author} {\bibnamefont {et~al.}},\ }\bibfield  {title} {\enquote {\bibinfo {title} {Preliminary results from the latin-american giant observatory space weather simulation chain},}\ }\href@noop {} {\bibfield  {journal} {\bibinfo  {journal} {Space Weather}\ }\textbf {\bibinfo {volume} {16}},\ \bibinfo {pages} {461--475} (\bibinfo {year} {2018})}\BibitemShut {NoStop}%
\bibitem [{\citenamefont {Asorey}, \citenamefont {Núñez},\ and\ \citenamefont {Suárez-Durán}(2018)}]{Asorey_etal2018}%
  \BibitemOpen
  \bibfield  {author} {\bibinfo {author} {\bibfnamefont {H.}~\bibnamefont {Asorey}}, \bibinfo {author} {\bibfnamefont {L.}~\bibnamefont {Núñez}}, \ and\ \bibinfo {author} {\bibfnamefont {M.}~\bibnamefont {Suárez-Durán}},\ }\bibfield  {title} {\enquote {\bibinfo {title} {Preliminary results from the latin american giant observatory space weather simulation chain},}\ }\href {\doibase 10.1002/2017SW001774} {\bibfield  {journal} {\bibinfo  {journal} {Space Weather}\ }\textbf {\bibinfo {volume} {16}},\ \bibinfo {pages} {461--475} (\bibinfo {year} {2018})},\ \Eprint {http://arxiv.org/abs/https://agupubs.onlinelibrary.wiley.com/doi/pdf/10.1002/2017SW001774} {https://agupubs.onlinelibrary.wiley.com/doi/pdf/10.1002/2017SW001774} \BibitemShut {NoStop}%
\bibitem [{\citenamefont {Heck}\ \emph {et~al.}(1998)\citenamefont {Heck}, \citenamefont {Kanapp}, \citenamefont {Capdevielle}, \citenamefont {Shatz},\ and\ \citenamefont {Thouw}}]{CORSIKA}%
  \BibitemOpen
  \bibfield  {author} {\bibinfo {author} {\bibfnamefont {D.}~\bibnamefont {Heck}}, \bibinfo {author} {\bibfnamefont {J.}~\bibnamefont {Kanapp}}, \bibinfo {author} {\bibfnamefont {N.}~\bibnamefont {Capdevielle}}, \bibinfo {author} {\bibfnamefont {G.}~\bibnamefont {Shatz}}, \ and\ \bibinfo {author} {\bibfnamefont {T.}~\bibnamefont {Thouw}},\ }\bibfield  {title} {\enquote {\bibinfo {title} {{CORSIKA:a MonteCarlo code to simulate extensive air showers,}},}\ }\href@noop {} {\bibfield  {journal} {\bibinfo  {journal} {FZKA}\ }\textbf {\bibinfo {volume} {6019, p. 1–90.}} (\bibinfo {year} {1998})}\BibitemShut {NoStop}%
\bibitem [{\citenamefont {Kneizys}\ \emph {et~al.}(1996)\citenamefont {Kneizys}, \citenamefont {Abreu}, \citenamefont {Anderson}, \citenamefont {Chetwynd} \emph {et~al.}}]{Kneizys1996}%
  \BibitemOpen
  \bibfield  {author} {\bibinfo {author} {\bibfnamefont {F.~X.}\ \bibnamefont {Kneizys}}, \bibinfo {author} {\bibfnamefont {L.~W.}\ \bibnamefont {Abreu}}, \bibinfo {author} {\bibfnamefont {G.~P.}\ \bibnamefont {Anderson}}, \bibinfo {author} {\bibfnamefont {J.~H.}\ \bibnamefont {Chetwynd}},  \emph {et~al.},\ }\href {http://web.gps.caltech.edu/{~}vijay/pdf/modrept.pdf} {\enquote {\bibinfo {title} {{The MODTRAN 2/3 report and LOWTRAN 7 model}},}\ }\bibinfo {type} {Tech. Rep.}\ (\bibinfo {year} {1996})\BibitemShut {NoStop}%
\bibitem [{\citenamefont {Grisales-Casadiegos}, \citenamefont {Sarmiento-Cano},\ and\ \citenamefont {N{\'u}{\~n}ez}(2022)}]{Grisalescasadiegos2022}%
  \BibitemOpen
  \bibfield  {author} {\bibinfo {author} {\bibfnamefont {J.}~\bibnamefont {Grisales-Casadiegos}}, \bibinfo {author} {\bibfnamefont {C.}~\bibnamefont {Sarmiento-Cano}}, \ and\ \bibinfo {author} {\bibfnamefont {L.}~\bibnamefont {N{\'u}{\~n}ez}},\ }\bibfield  {title} {\enquote {\bibinfo {title} {Impact of global data assimilation system atmospheric models on astroparticle showers},}\ }\href@noop {} {\bibfield  {journal} {\bibinfo  {journal} {Canadian Journal of Physics}\ }\textbf {\bibinfo {volume} {100}},\ \bibinfo {pages} {152--157} (\bibinfo {year} {2022})}\BibitemShut {NoStop}%
\bibitem [{\citenamefont {V{\'a}squez-Ram{\'\i}rez}\ \emph {et~al.}(2020)\citenamefont {V{\'a}squez-Ram{\'\i}rez}, \citenamefont {Su{\'a}rez-Dur{\'a}n}, \citenamefont {Jaimes-Motta}, \citenamefont {Calder{\'o}n-Ardila}, \citenamefont {Pe{\~n}a-Rodr{\'\i}guez}, \citenamefont {S{\'a}nchez-Villafrades}, \citenamefont {Sanabria-G{\'o}mez}, \citenamefont {Asorey},\ and\ \citenamefont {N{\'u}{\~n}ez}}]{VasquezEtal2020}%
  \BibitemOpen
  \bibfield  {author} {\bibinfo {author} {\bibfnamefont {A.}~\bibnamefont {V{\'a}squez-Ram{\'\i}rez}}, \bibinfo {author} {\bibfnamefont {M.}~\bibnamefont {Su{\'a}rez-Dur{\'a}n}}, \bibinfo {author} {\bibfnamefont {A.}~\bibnamefont {Jaimes-Motta}}, \bibinfo {author} {\bibfnamefont {R.}~\bibnamefont {Calder{\'o}n-Ardila}}, \bibinfo {author} {\bibfnamefont {J.}~\bibnamefont {Pe{\~n}a-Rodr{\'\i}guez}}, \bibinfo {author} {\bibfnamefont {J.}~\bibnamefont {S{\'a}nchez-Villafrades}}, \bibinfo {author} {\bibfnamefont {J.}~\bibnamefont {Sanabria-G{\'o}mez}}, \bibinfo {author} {\bibfnamefont {H.}~\bibnamefont {Asorey}}, \ and\ \bibinfo {author} {\bibfnamefont {L.}~\bibnamefont {N{\'u}{\~n}ez}},\ }\bibfield  {title} {\enquote {\bibinfo {title} {Simulated response of mute, a hybrid muon telescope},}\ }\href@noop {} {\bibfield  {journal} {\bibinfo  {journal} {Journal of Instrumentation}\ }\textbf {\bibinfo {volume} {15}},\ \bibinfo {pages} {P08004} (\bibinfo {year} {2020})}\BibitemShut {NoStop}%
\bibitem [{\citenamefont {Tishevsky}\ \emph {et~al.}(2024{\natexlab{b}})\citenamefont {Tishevsky}, \citenamefont {Dubinin}, \citenamefont {Isupov}, \citenamefont {Ladygin}, \citenamefont {Nigmatkulov}, \citenamefont {Reznikov}, \citenamefont {Teterin}, \citenamefont {Volkov}, \citenamefont {Zakharov},\ and\ \citenamefont {Zhurkina}}]{Tishevsky2024}%
  \BibitemOpen
  \bibfield  {author} {\bibinfo {author} {\bibfnamefont {A.~V.}\ \bibnamefont {Tishevsky}}, \bibinfo {author} {\bibfnamefont {F.~A.}\ \bibnamefont {Dubinin}}, \bibinfo {author} {\bibfnamefont {A.~Y.}\ \bibnamefont {Isupov}}, \bibinfo {author} {\bibfnamefont {V.~P.}\ \bibnamefont {Ladygin}}, \bibinfo {author} {\bibfnamefont {G.~A.}\ \bibnamefont {Nigmatkulov}}, \bibinfo {author} {\bibfnamefont {S.~G.}\ \bibnamefont {Reznikov}}, \bibinfo {author} {\bibfnamefont {P.~E.}\ \bibnamefont {Teterin}}, \bibinfo {author} {\bibfnamefont {I.~S.}\ \bibnamefont {Volkov}}, \bibinfo {author} {\bibfnamefont {A.~M.}\ \bibnamefont {Zakharov}}, \ and\ \bibinfo {author} {\bibfnamefont {A.~O.}\ \bibnamefont {Zhurkina}},\ }\bibfield  {title} {\enquote {\bibinfo {title} {Development of the spd beam–beam counter scintillation detector prototype with fers-5200 front-end readout system},}\ }\href {\doibase 10.1134/S1063778824700510} {\bibfield  {journal} {\bibinfo  {journal} {Physics of Atomic Nuclei}\ }\textbf {\bibinfo {volume}
  {87}},\ \bibinfo {pages} {451–458} (\bibinfo {year} {2024}{\natexlab{b}})}\BibitemShut {NoStop}%
\bibitem [{\citenamefont {Zakharov}\ \emph {et~al.}(2024)\citenamefont {Zakharov}, \citenamefont {Dubinin}, \citenamefont {Isupov}, \citenamefont {Ladygin}, \citenamefont {Manakonov}, \citenamefont {Nigmatkulov}, \citenamefont {Reznikov}, \citenamefont {Teterin}, \citenamefont {Tishevsky}, \citenamefont {Volkov},\ and\ \citenamefont {Zhurkina}}]{Zakharov2024}%
  \BibitemOpen
  \bibfield  {author} {\bibinfo {author} {\bibfnamefont {A.~M.}\ \bibnamefont {Zakharov}}, \bibinfo {author} {\bibfnamefont {F.~A.}\ \bibnamefont {Dubinin}}, \bibinfo {author} {\bibfnamefont {A.~Y.}\ \bibnamefont {Isupov}}, \bibinfo {author} {\bibfnamefont {V.~P.}\ \bibnamefont {Ladygin}}, \bibinfo {author} {\bibfnamefont {A.~D.}\ \bibnamefont {Manakonov}}, \bibinfo {author} {\bibfnamefont {G.~A.}\ \bibnamefont {Nigmatkulov}}, \bibinfo {author} {\bibfnamefont {S.~G.}\ \bibnamefont {Reznikov}}, \bibinfo {author} {\bibfnamefont {P.~E.}\ \bibnamefont {Teterin}}, \bibinfo {author} {\bibfnamefont {A.~V.}\ \bibnamefont {Tishevsky}}, \bibinfo {author} {\bibfnamefont {I.~S.}\ \bibnamefont {Volkov}}, \ and\ \bibinfo {author} {\bibfnamefont {A.~O.}\ \bibnamefont {Zhurkina}},\ }\bibfield  {title} {\enquote {\bibinfo {title} {Tile detector configurations testing for the spd beam-beam counter prototype},}\ }\href {\doibase 10.1134/S1547477124701218} {\bibfield  {journal} {\bibinfo  {journal} {Physics of Particles and
  Nuclei Letters}\ }\textbf {\bibinfo {volume} {21}},\ \bibinfo {pages} {735–738} (\bibinfo {year} {2024})}\BibitemShut {NoStop}%
\bibitem [{\citenamefont {{CAEN S.p.A.}}(2024{\natexlab{a}})}]{FERS5200}%
  \BibitemOpen
  \bibfield  {author} {\bibinfo {author} {\bibnamefont {{CAEN S.p.A.}}},\ }\href@noop {} {\emph {\bibinfo {title} {{FERS-5200 - Front-End Readout System}}}} (\bibinfo {year} {2024}{\natexlab{a}}),\ \bibinfo {note} {{Available: \url{https://www.caen.it/products/fers-5200}}}\BibitemShut {NoStop}%
\bibitem [{\citenamefont {{CAEN S.p.A.}}(2023)}]{UM7945}%
  \BibitemOpen
  \bibfield  {author} {\bibinfo {author} {\bibnamefont {{CAEN S.p.A.}}},\ }\href@noop {} {\emph {\bibinfo {title} {{UM7945 - A5202/DT5202 User Manual}}}} (\bibinfo {year} {2023}),\ \bibinfo {note} {{Available: \url{https://www.caen.it/products/um7945}}}\BibitemShut {NoStop}%
\bibitem [{\citenamefont {Venaruzzo}\ \emph {et~al.}(2020)\citenamefont {Venaruzzo}, \citenamefont {Abba}, \citenamefont {Tintori},\ and\ \citenamefont {Venturini}}]{Venaruzzo2020}%
  \BibitemOpen
  \bibfield  {author} {\bibinfo {author} {\bibfnamefont {M.}~\bibnamefont {Venaruzzo}}, \bibinfo {author} {\bibfnamefont {A.}~\bibnamefont {Abba}}, \bibinfo {author} {\bibfnamefont {C.}~\bibnamefont {Tintori}}, \ and\ \bibinfo {author} {\bibfnamefont {Y.}~\bibnamefont {Venturini}},\ }\bibfield  {title} {\enquote {\bibinfo {title} {Fers-5200: a distributed front-end readout system for multidetector arrays},}\ }\href {https://arxiv.org/abs/2010.15688} {\bibfield  {journal} {\bibinfo  {journal} {IEEE Transactions on Nuclear Science}\ }\textbf {\bibinfo {volume} {XX}},\ \bibinfo {pages} {1--8} (\bibinfo {year} {2020})},\ \Eprint {http://arxiv.org/abs/2010.15688} {arXiv:2010.15688 [physics.ins-det]} \BibitemShut {NoStop}%
\bibitem [{\citenamefont {{CAEN S.p.A.}}(2024{\natexlab{b}})}]{UM7946}%
  \BibitemOpen
  \bibfield  {author} {\bibinfo {author} {\bibnamefont {{CAEN S.p.A.}}},\ }\href@noop {} {\emph {\bibinfo {title} {{UM7946 - Janus 5202 Software User Manual}}}} (\bibinfo {year} {2024}{\natexlab{b}}),\ \bibinfo {note} {{Available: \url{https://www.caen.it/products/um7946}}}\BibitemShut {NoStop}%
\end{thebibliography}%

\end{document}